\documentclass{aastex62}
\usepackage{amsmath}
\usepackage{amssymb}
\usepackage{graphicx}
\usepackage{longtable}
\usepackage{xcolor}

\newcommand{\tess}{\textit{TESS}}
\newcommand{\zaspe}{\texttt{ZASPE}}
\newcommand{\species}{\texttt{SPECIES}}
\newcommand{\ariadne}{\texttt{ARIADNE}}
\newcommand{\spc}{\texttt{SPC}}
\newcommand{\moog}{\texttt{MOOG}}
\newcommand{\exonailer}{\texttt{juliet}}
\newcommand{\emperor}{\texttt{EMPEROR}}
\newcommand{\exofast}{\texttt{EXOFASTv2}}

\newcommand{\rptess}{4.72~$\pm$~0.23~$R_{\oplus}$}
\newcommand{\mplan}{29.32$^{+0.78}_{-0.81}$~$M_{\oplus}$}

\newcommand{\perdays}{0.792054~$\pm$~0.000014~d}

\newcommand{\starname}{LTT\,9779}
\newcommand{\planet}{LTT\,9779~b}
\newcommand{\tessdepth}{1584~$\pm$~43~ppm}
\newcommand{\msun}{$M_{\odot}$}
\newcommand{\rsun}{$R_{\odot}$}
\newcommand{\me}{$M_{\oplus}$}
\newcommand{\re}{$R_{\oplus}$}
\newcommand{\mj}{$M_{\rm{J}}$}

\newcommand{\kms}{km\,s$^{-1}$}
\newcommand{\vsini}{\ensuremath{v \sin{i}}}
\newcommand{\feh}{\ensuremath{{\rm [Fe/H]}}}
\newcommand{\meh}{\ensuremath{{\rm [m/H]}}}
\newcommand{\teff}{\ensuremath{T_{\rm eff}}}
\newcommand{\rstar}{\ensuremath{{\rm R}_{\star}}}
\newcommand{\mstar}{\ensuremath{{\rm M}_{\star}}}
\newcommand{\logg}{\ensuremath{\log{g}}}
\newcommand{\logrhk}{\ensuremath{logR'_{HK,HARPS}}}
\newcommand{\mpl}{\ensuremath{{\rm M_p}}}
\newcommand{\rpl}{\ensuremath{{\rm R_p}}}
\newcommand{\teq}{\ensuremath{T_{\rm eq}}}
\newcommand{\rhopl}{\ensuremath{{\rm \rho_p}}}

\begin{document}

\title{An Ultra-Hot Neptune in the Neptune desert}
%% Notice placement of commas and superscripts and use of & in the author list

\author{James S.\ Jenkins}
\altaffiliation{Corresponding author E-mail: \texttt{jjenkins@das.uchile.cl}}
\affil{Departamento de Astronomía, Universidad de Chile, Camino El Observatorio 1515, Las Condes, Santiago, Chile }
\affil{Centro de Astrof\'isica y Tecnolog\'ias Afines (CATA), Casilla 36-D, Santiago, Chile}

\author{Mat\'ias R. D\'iaz}
\affil{Departamento de Astronomía, Universidad de Chile, Camino El Observatorio 1515, Las Condes, Santiago, Chile }
\affil{Centro de Astrof\'isica y Tecnolog\'ias Afines (CATA), Casilla 36-D, Santiago, Chile}

\author{Nicol\'as T. Kurtovic}
\affil{Departamento de Astronomía, Universidad de Chile, Camino El
  Observatorio 1515, Las Condes, Santiago, Chile }

\author{N\'estor Espinoza}
\affil{Space Telescope Science Institute, 3700 San Martin Drive, Baltimore, MD 21218, USA}

\author{Jose I. Vines}
\affil{Departamento de Astronomía, Universidad de Chile, Camino El
  Observatorio 1515, Las Condes, Santiago, Chile }

\author{Pablo A. Pe\~na Rojas}
\affil{Departamento de Astronomía, Universidad de Chile, Camino El
  Observatorio 1515, Las Condes, Santiago, Chile }

\author{Rafael Brahm}
\affil{Facultad de Ingenier\'ia y Ciencias, Universidad Adolfo Ib\'a\~nez, Av.\ Diagonal las Torres 2640, Pe\~nalol\'en, Santiago, Chile}
\affil{Millennium Institute for Astrophysics, Av.\ Vicu\~{n}a Mackenna 4860, 782-0436 Macul, Santiago, Chile}

\author{Pascal Torres}
\affil{Center of Astro-Engineering UC, Pontificia Universidad
  Cat\'olica de Chile, Av.  Vicu\~na Mackenna 4860, 7820436 Macul,
  Santiago, Chile}

\author{P\'ia Cort\'es-Zuleta}
\affil{Departamento de Astronomía, Universidad de Chile, Camino El
  Observatorio 1515, Las Condes, Santiago, Chile }

\author{Maritza G. Soto}
\affil{School of Physics and Astronomy, Queen Mary University of London, 327 Mile End Road, London E1 4NS, UK}

\author{Eric D. Lopez}
\affil{NASA Goddard Space Flight Center, 8800 Greenbelt Rd, Greenbelt, MD 20771, USA}

\author{George W. King}
\affil{Department of Physics, University of Warwick, Gibbet Hill Road, Coventry CV4 7AL, UK}
\affil{Centre for Exoplanets and Habitability, University of Warwick, Gibbet Hill Road, Coventry CV4 7AL, UK}

\author{Peter J. Wheatley}
\affil{Department of Physics, University of Warwick, Gibbet Hill Road, Coventry CV4 7AL, UK}
\affil{Centre for Exoplanets and Habitability, University of Warwick, Gibbet Hill Road, Coventry CV4 7AL, UK}

\author{Joshua N. Winn}
\affil{Department of Astrophysical Sciences, Princeton University, 4 Ivy Lane, Princeton, NJ 08544 USA}

\author{David R. Ciardi}
\affil{NASA Exoplanet Science Institute/Caltech Pasadena, CA, USA}

\author{George Ricker}
\affil{Department of Physics and Kavli Institute for Astrophysics and Space Research, Massachusetts Institute of Technology, Cambridge, MA 02139, USA}

\author{Roland Vanderspek}
\affil{Royal Observatory of Belgium, Ringlaan 3, B-1180 Brussel, Belgium}

\author{David W. Latham}
\affil{Center for Astrophysics ${\rm \mid}$ Harvard {\rm \&} Smithsonian, 60 Garden Street, Cambridge, MA 02138, USA}

\author{Sara Seager}
\affil{Department of Physics and Kavli Institute for Astrophysics and Space Research, Massachusetts Institute of Technology, Cambridge, MA 02139, USA}
\affil{Department of Earth and Planetary Sciences, Massachusetts Institute of Technology, Cambridge, MA 02139, USA}

\author{ Jon M. Jenkins}
\affil{NASA Ames Research Center, Moffett Field, CA, 94035}

\author{Charles A. Beichman}
\affil{NASA Exoplanet Science Institute/Caltech Pasadena, CA, USA}

\author{Allyson Bieryla}
\affil{Center for Astrophysics ${\rm \mid}$ Harvard {\rm \&} Smithsonian, 60 Garden Street, Cambridge, MA 02138, USA}

\author{Christopher J. Burke}
\affil{Department of Physics and Kavli Institute for Astrophysics and Space Research, Massachusetts Institute of Technology, Cambridge, MA 02139, USA}

\author{Jessie L. Christiansen}
\affil{NASA Exoplanet Science Institute/Caltech Pasadena, CA, USA}

\author{Christopher E. Henze}
\affil{NASA Ames Research Center, Moffett Field, CA, 94035}

\author{Todd C. Klaus}
\affil{NASA Ames Research Center, Moffett Field, CA, 94035}

\author{Sean McCauliff}
\affil{NASA Ames Research Center, Moffett Field, CA, 94035}

\author{Mayuko Mori}
\affil{Department of Astronomy, The University of Tokyo, 7-3-1 Hongo, Bunkyo-ku, Tokyo 113-0033, Japan}

\author{Norio Narita}
\affil{Komaba Institute for Science, The University of Tokyo, 3-8-1 Komaba, Meguro, Tokyo 153-8902, Japan}
\affil{JST, PRESTO, 3-8-1 Komaba, Meguro, Tokyo 153-8902, Japan}
\affil{Astrobiology Center, 2-21-1 Osawa, Mitaka, Tokyo 181-8588, Japan}
\affil{National Astronomical Observatory of Japan, 2-21-1 Osawa, Mitaka, Tokyo 181-8588, Japan}
\affil{Instituto de Astrof\'isica de Canarias (IAC), 38205 La Laguna, Tenerife, Spain}

\author{Taku Nishiumi}
\affil{Department of Physics, Kyoto Sangyo University, Motoyama, Kamigamo, Kita-ku, Kyoto, 603-8555 Japan}

\author{Motohide Tamura}
\affil{Department of Astronomy, The University of Tokyo, 7-3-1 Hongo, Bunkyo-ku, Tokyo 113-0033, Japan}
\affil{Astrobiology Center, 2-21-1 Osawa, Mitaka, Tokyo 181-8588, Japan}
\affil{National Astronomical Observatory of Japan, 2-21-1 Osawa, Mitaka, Tokyo 181-8588, Japan}

\author{Jerome Pitogo de Leon}
\affil{Department of Astronomy, The University of Tokyo, 7-3-1 Hongo, Bunkyo-ku, Tokyo 113-0033, Japan}

\author{Samuel N. Quinn}
\affil{Center for Astrophysics ${\rm \mid}$ Harvard {\rm \&} Smithsonian, 60 Garden Street, Cambridge, MA 02138, USA}

\author{Jesus Noel Villase\~nor}
\affil{Department of Physics and Kavli Institute for Astrophysics and Space Research, Massachusetts Institute of Technology, Cambridge, MA 02139, USA}

\author{Michael Vezie}
\affil{Department of Physics and Kavli Institute for Astrophysics and Space Research, Massachusetts Institute of Technology, Cambridge, MA 02139, USA}

\author{Jack J. Lissauer}
\affil{NASA Ames Research Center, Moffett Field, CA, 94035}

\author{Karen A.\ Collins}
\affil{Center for Astrophysics ${\rm \mid}$ Harvard {\rm \&} Smithsonian, 60 Garden Street, Cambridge, MA 02138, USA}

\author{Kevin I.\ Collins}
\affil{George Mason University, 4400 University Drive, Fairfax, VA, 22030 USA}

\author{Giovanni Isopi}
\affil{Campo Catino Astronomical Observatory, Regione Lazio, Guarcino (FR), 03010 Italy}

\author{Franco Mallia}
\affil{Campo Catino Astronomical Observatory, Regione Lazio, Guarcino (FR), 03010 Italy}

\author{Andrea Ercolino}
\affil{Campo Catino Astronomical Observatory, Regione Lazio, Guarcino (FR), 03010 Italy}

\author{Cristobal Petrovich}
\affil{Canadian Institute for Theoretical Astrophysics, University of Toronto, 60 St George Street, ON M5S 3H8, Canada}
\affil{Centre for Planetary Sciences, Department of Physical \& Environmental Sciences, University of Toronto at Scarborough, Toronto, Ontario M1C 1A4, Canada}

\author{Andr\'es Jord\'an}
\affil{Facultad de Ingenier\'ia y Ciencias, Universidad Adolfo Ib\'a\~nez, Av.\ Diagonal las Torres 2640, Pe\~nalol\'en, Santiago, Chile}
\affil{Millennium Institute for Astrophysics, Av.\ Vicu\~{n}a Mackenna 4860, 782-0436 Macul, Santiago, Chile}

\author{Jack~S. Acton}
\affil{Department of Physics and Astronomy, University of Leicester, University Road, Leicester, LE1 7RH, UK}

\author{David J. Armstrong}
\affil{Department of Physics, University of Warwick, Gibbet Hill Road, Coventry CV4 7AL, UK}
\affil{Centre for Exoplanets and Habitability, University of Warwick, Gibbet Hill Road, Coventry CV4 7AL, UK}

\author{Daniel Bayliss}
\affil{Department of Physics, University of Warwick, Gibbet Hill Road, Coventry CV4 7AL, UK}

\author{Fran\c{c}ois Bouchy}
\affil{Observatoire de Gen{\`e}ve, Universit{\'e} de Gen{\`e}ve, 51 Ch. des Maillettes, 1290 Sauverny, Switzerland}

\author{Claudia~Belardi}
\affil{Department of Physics and Astronomy, University of Leicester, University Road, Leicester, LE1 7RH, UK}

\author{Edward M. Bryant}
\affil{Department of Physics, University of Warwick, Gibbet Hill Road, Coventry CV4 7AL, UK}
\affil{Centre for Exoplanets and Habitability, University of Warwick, Gibbet Hill Road, Coventry CV4 7AL, UK}

\author{Matthew R. Burleigh}
\affil{Department of Physics and Astronomy, University of Leicester, University Road, Leicester, LE1 7RH, UK}

\author{Juan Cabrera}
\affil{Institute of Planetary Research, German Aerospace Center, Rutherfordstrasse 2, 12489 Berlin, Germany}

\author{Sarah L. Casewell}
\affil{Department of Physics and Astronomy, University of Leicester, University Road, Leicester, LE1 7RH, UK}

\author{Alexander Chaushev}
\affil{Center for Astronomy and Astrophysics, TU Berlin, Hardenbergstr. 36, D-10623 Berlin, Germany}

\author{Benjamin F. Cooke}
\affil{Department of Physics, University of Warwick, Gibbet Hill Road, Coventry CV4 7AL, UK}
\affil{Centre for Exoplanets and Habitability, University of Warwick, Gibbet Hill Road, Coventry CV4 7AL, UK}

\author{Philipp Eigm\"uller}
\affil{Institute of Planetary Research, German Aerospace Center, Rutherfordstrasse 2, 12489 Berlin, Germany}

\author{Anders Erikson}
\affil{Institute of Planetary Research, German Aerospace Center, Rutherfordstrasse 2, 12489 Berlin, Germany}

\author{Emma Foxell}
\affil{Department of Physics, University of Warwick, Gibbet Hill Road, Coventry CV4 7AL, UK}
\affil{Centre for Exoplanets and Habitability, University of Warwick, Gibbet Hill Road, Coventry CV4 7AL, UK}

\author{Boris T. G\"ansicke}
\affil{Department of Physics, University of Warwick, Gibbet Hill Road, Coventry CV4 7AL, UK}

\author{Samuel Gill}
\affil{Department of Physics, University of Warwick, Gibbet Hill Road, Coventry CV4 7AL, UK}
\affil{Centre for Exoplanets and Habitability, University of Warwick, Gibbet Hill Road, Coventry CV4 7AL, UK}

\author{Edward Gillen}
\altaffiliation{Winton Fellow}
\affil{Astrophysics Group, Cavendish Laboratory, J.J. Thomson Avenue, Cambridge CB3 0HE, UK}

\author{Maximilian N. G\"unther}
\affil{Department of Physics and Kavli Institute for Astrophysics and Space Research, Massachusetts Institute of Technology, Cambridge, MA 02139, USA}

\author{Michael R. Goad}
\affil{Department of Physics and Astronomy, University of Leicester, University Road, Leicester, LE1 7RH, UK}

\author{Matthew J. Hooton}
\affil{Astrophysics Research Centre, School of Mathematics and Physics, Queen's University Belfast, BT7 1NN Belfast, UK}

\author{James A. G. Jackman}
\affil{Department of Physics, University of Warwick, Gibbet Hill Road, Coventry CV4 7AL, UK}
\affil{Centre for Exoplanets and Habitability, University of Warwick, Gibbet Hill Road, Coventry CV4 7AL, UK}

\author{Tom Louden}
\affil{Department of Physics, University of Warwick, Gibbet Hill Road, Coventry CV4 7AL, UK}
\affil{Centre for Exoplanets and Habitability, University of Warwick, Gibbet Hill Road, Coventry CV4 7AL, UK}

\author{James McCormac}
\affil{Department of Physics, University of Warwick, Gibbet Hill Road, Coventry CV4 7AL, UK}
\affil{Centre for Exoplanets and Habitability, University of Warwick, Gibbet Hill Road, Coventry CV4 7AL, UK}

\author{Maximiliano Moyano}
\affil{Instituto de Astronom\'ia, Universidad Cat\'olica del Norte, Angamos 0610, 1270709, Antofagasta, Chile}

\author{Louise D. Nielsen}
\affil{Observatoire de Gen{\`e}ve, Universit{\'e} de Gen{\`e}ve, 51 Ch. des Maillettes, 1290 Sauverny, Switzerland}

\author{Don Pollacco}
\affil{Department of Physics, University of Warwick, Gibbet Hill Road, Coventry CV4 7AL, UK}
\affil{Centre for Exoplanets and Habitability, University of Warwick, Gibbet Hill Road, Coventry CV4 7AL, UK}

\author{Didier Queloz}
\affil{Astrophysics Group, Cavendish Laboratory, J.J. Thomson Avenue, Cambridge CB3 0HE, UK}

\author{Heike Rauer}
\affil{Institute of Planetary Research, German Aerospace Center, Rutherfordstrasse 2, 12489 Berlin, Germany}
\affil{Center for Astronomy and Astrophysics, TU Berlin, Hardenbergstr. 36, D-10623 Berlin, Germany}
\affil{Institute of Geological Sciences, FU Berlin, Malteserstr. 74-100, D-12249 Berlin, Germany}

\author{Liam Raynard}
\affil{Department of Physics and Astronomy, University of Leicester, University Road, Leicester, LE1 7RH, UK}

\author{Alexis M. S. Smith}
\affil{Institute of Planetary Research, German Aerospace Center, Rutherfordstrasse 2, 12489 Berlin, Germany}

\author{Rosanna H. Tilbrook}
\affil{Department of Physics and Astronomy, University of Leicester, University Road, Leicester, LE1 7RH, UK}

\author{Ruth Titz-Weider}
\affil{Institute of Planetary Research, German Aerospace Center, Rutherfordstrasse 2, 12489 Berlin, Germany}

\author{Oliver Turner}
\affil{Observatoire de Gen{\`e}ve, Universit{\'e} de Gen{\`e}ve, 51 Ch. des Maillettes, 1290 Sauverny, Switzerland}

\author{St\'{e}phane Udry}
\affil{Observatoire de Gen{\`e}ve, Universit{\'e} de Gen{\`e}ve, 51 Ch. des Maillettes, 1290 Sauverny, Switzerland}

\author{Simon. R. Walker}
\affil{Department of Physics, University of Warwick, Gibbet Hill Road, Coventry CV4 7AL, UK}

\author{Christopher A. Watson}
\affil{Astrophysics Research Centre, School of Mathematics and Physics, Queen's University Belfast, BT7 1NN Belfast, UK}

\author{Richard G. West}
\affil{Department of Physics, University of Warwick, Gibbet Hill Road, Coventry CV4 7AL, UK}
\affil{Centre for Exoplanets and Habitability, University of Warwick, Gibbet Hill Road, Coventry CV4 7AL, UK}

\author{Enric Palle}
\affil{Instituto de Astrof\'isica de Canarias (IAC), 38205 La Laguna, Tenerife, Spain}
\affil{Departamento  de  Astrof\'isica,  Universidad  de  La  Laguna (ULL), 38206, La Laguna, Tenerife, Spain}

\author{Carl Ziegler}
\affil{Dunlap Institute for Astronomy and Astrophysics, University of Toronto, 50 St. George Street, Toronto, Ontario M5S 3H4, Canada}

\author{Nicholas Law}
\affil{Department of Physics and Astronomy, The University of North Carolina at Chapel Hill, Chapel Hill, NC 27599-3255, USA}

\author{Andrew W. Mann}
\affil{Department of Physics and Astronomy, The University of North Carolina at Chapel Hill, Chapel Hill, NC 27599-3255, USA}

\begin{abstract}
About one out of 200 Sun-like stars has a planet with an orbital period
shorter than one day: an ultra-short-period planet \citep{sanchis-ojeda14,winn18}.
All of the previously known ultra-short-period planets are either hot Jupiters, with sizes
above 10~Earth radii (\re), or apparently rocky planets
smaller than 2~\re. Such lack of planets of intermediate size (the ``hot Neptune desert'') has been interpreted as the inability of low-mass planets to retain any
hydrogen/helium (H/He) envelope in the face of strong stellar irradiation.
Here, we report the discovery of an ultra-short-period planet with a radius of
4.6~\re\ and a mass of 29~\me, firmly in the hot Neptune desert.
Data from the Transiting Exoplanet Survey Satellite \citep{ricker15} revealed transits of the bright Sun-like star \starname\, every 0.79 days.
The planet's mean density is similar to that of Neptune, and according
to thermal evolution models, it has a H/He-rich envelope constituting 9.0$^{+2.7}_{-2.9}$\% of the total mass.
With an equilibrium temperature around 2000\,K, it is unclear how this 
``ultra-hot Neptune'' managed to retain such an envelope.
Follow-up observations of the planet's atmosphere to better understand its origin and physical nature will be facilitated by the star's brightness ($V_{\rm mag}=9.8$).
\end{abstract}

\section{Main Manuscript}

Using high precision photometry from Sector 2 of the Transiting
Exoplanet Survey Satellite (\tess) mission at a cadence of two
minutes, a candidate transiting planet was flagged for the star
\starname\, \citep{jenkins16_spie}.  The candidate was released as a \tess\, Alert in October 2018, and assigned the TESS Object of Interest (TOI) tagname TOI-193 (TIC183985250).  The \tess\, lightcurve was scrutinised prior to its public release. No transit depth variations were apparent, no motion of the stellar image was detected during transits, 
and no secondary eclipses could be found.
Data from the Gaia spacecraft \citep{gaia16,gaia18}
revealed only one background star within the \tess\ photometric
aperture, but it is 5~mag fainter than \starname\ and hence cannot
be the source of the transit-like signals, and no significant excess
scatter was witnessed in the Gaia measurements.
The lack of all these abnormalities supported the initial interpretation
that the transit signals are due to a planet with an orbital period of 19~hours and a radius of 3.96~\re.  

We also observed four complete transits with ground-based facilities:
three with the Las Cumbres Observatory (LCO) and one with the Next
Generation Transit Survey (NGTS; \citealp{wheatley18}) telescopes.
The LCO and NGTS data have a similar precision to the \tess\, light curve and much
better angular resolution.
The observed transit depths were in agreement with the depth observed
with \tess. High-angular resolution imaging of LTT 9779 was
  performed with adaptive optics in the near-infrared using NIRC2 at
  the Keck Observatory, and with speckle imaging in the optical using
  HRCam on SOAR at the Cerro-Tololo Inter-American Observatory.  No companions were detected within a radius of $3''$ down to a contrast level of 7.5 magnitudes, and no bright close binary was seen with a resolution of $0.05''$ (see the SI).
These observations sharply reduce the possibility that an unresolved background star is the source of the transits.
We also tested the probability of having background or foreground
stars within a region of $0.1''$ separation (AO limit) from the star,
using a Besan\c{c}on \citep{robin03} model of the galaxy. The model indicates we can expect over 2200 stars in a 1 square degree field around \starname\, providing a probability of only 0.0005\% of having a star down to a magnitude limit of 21 in $V$ contaminating the lightcurves. If we consider only objects bright enough to cause contamination of the transit depth that would significantly alter the planet properties, this probability drops even more (see the Supplementary Information (SI) for more details).  Furthermore, although there exists a 13.5\% probability that \starname\, could be part of a binary system that passes within this separation limit, spectral analysis rules out all allowable masses whose contaminant light that would be required to push \planet\, outside of the Neptune desert.

Final confirmation of the planet's existence came from high-cadence
radial-velocity observations with the High Accuracy Radial-velocity
Planet Searcher (HARPS; \citealp{pepe00}).  A sinusoidal
radial-velocity (RV) signal was detected with the \emperor\ code \citep{JenkinsEtal2018} independently of the transit data, but with a matching
orbital period and phase. 
No other significant signals were detected, nor were any longer-term
trends, ruling out additional massive planets with orbital periods of a few years or less.
Likewise, no transit-timing variations were detected (see the SI).

To determine the stellar properties, we combined the Gaia data with
spectral information
from HARPS, along with other spectra from the Tillinghast Reflector
Echelle Spectrograph (TRES; \citealp{furesz08}) and the Network of
Robotic Echelle Spectrographs (NRES; \citealp{siverd18}) and compare the star's observable properties to the
outputs from theoretical stellar-evolutionary models (MIST and Y2).  
We also used our new \ariadne\, code to precisely calculate the effective temperature and stellar radius (see SI for more information on these methods).  
The star was found to have a mass, radius, and age of 1.02$^{+0.02}_{-0.03}$~\msun, 0.949$\pm$0.006~\rsun, and 2.0$^{+1.3}_{-0.9}$~Gyrs, respectively.  The effective temperature and surface gravity are consistent with a main-sequence star 
slightly cooler than the Sun.  The spectra also revealed the star to be
approximately twice as metal-rich as the Sun (\feh\, = +0.25$\pm$0.04~dex). Table~\ref{tab:star} displays all the parameter values.

We utilised the \exonailer\ code \citep{juliet} to perform a joint analysis of
the transit and radial-velocity data (Figure~1).  
The period, mass, and radius of the planet were found to be
\perdays, \mplan, and \rptess, respectively.  The orbit is circular to
within the limits allowed by the radial-velocity data (the posterior odds ratio is 49:1 in favor of a circular model over an eccentric model).

\planet\, sits in the hot Neptune desert \citep{mazeh16}
(Figure~2), providing an opportunity to study the link between short-period gas giants and lower mass super-Earths.
The planet's mean density is similar to that of Neptune, and the planet's mass
and radius are incompatible with either a pure rock or pure water
composition (Figure~3), implying that it possesses a substantial H/He
gaseous atmosphere.  Using 1-D thermal evolution models from
\citet{lopez14}, assuming a silicate and iron core and a solar
composition gaseous envelope, we find a planet core mass of
27.9$^{+1.2}_{-1.0}$~\me, and an atmospheric mass fraction
of 9.0$^{+2.7}_{-2.9}$\%. We also tested other planet structures,
and even in the limiting case of a non-physical pure water-world, there still
exists a significant H/He-rich envelope, at the level of 2.2$^{+1.1}_{-1.6}$\%.  When combined with the high equilibrium temperature for the planet of 1978~$\pm$~19 K, this makes \planet\, an excellent target for future transmission spectroscopy, secondary eclipse studies, and phase variation analyses.  All of the planetary model parameters are in Table~\ref{tab:planet}. 

\planet\, is the most highly irradiated Neptune-sized planet yet
found. It is firmly in the region of parameter space known as the
"evaporation desert" where observations have shown a clear absence of
similarly sized planets \citep{sanchis-ojeda14,Lundkvist2016}, and
models of photo-evaporative atmospheric escape predict that such low
density gaseous atmospheres should be evaporated on short timescales
\citep{lopez17,owen17}. As \planet\, is a mature planet found in this
desert, it is a particularly high priority target for transmission
spectroscopy at wavelengths that probe low density material escaping
from planetary upper atmospheres such as Lyman Alpha
\citep{ehrenreich15}, FUV metal-lines \citep{VidalMadjar2004}, Ca and
Fe lines \citep{casasayas-barris19}, and the 1.083 $\mu$m Helium line \citep{nortmann18}.

An interesting comparison can be made between \planet\ and NGTS-4~b \citep{west18},
the most similar of all the other known planets.  NGTS-4~b is not as
hot ($\langle$\teq\, $\rangle^{a} =1650\pm400$ K) or
short-period ($P =1.337351\pm0.000008$~d) as \planet, has a much higher density of 3.45~g/cm$^3$, and
orbits a metal-poor star ($[M/H]=-0.28\pm0.10$~dex).  These characteristics may be clues
that the two planets formed differently: NGTS-4~b may have formed as a
relatively small and dense world, whereas \planet\, started life as a
much larger and less dense planet (see Figure~4).  Indeed,
photoevaporation models posit that the bulk population of ultra-short period planets
form by growing to around 3~\me, through the accretion of various
amounts of light elements from the proto-planetary disk.  The intense
radiation from the young star then evaporates these close-in planets
over an interval on the order of 10$^8$~yrs, leaving behind small
rocky planets with radii less than 1.5~\re\, \citep{owen17}.  The more
massive population of planets can hold onto the bulk of their
envelopes until the star becomes quiescent, leaving behind planets
with radii 2$\--$3~\re.  However, these planets are generally found to
have orbital periods beyond one day, similar to NGTS-4~b, reaching out
to 100~days or so.  Ultra-short period planets with these radii are rare, and it may
be that since \planet\, likely has a large mass, it can hold onto a
high fraction of its atmosphere. 
It could also have migrated to its current position over a
longer dynamical timescale, 10$^9$~years, 
not leaving enough time to blow-off a large fraction of its atmosphere by photoevaporation.  

  Assuming energy-limited atmospheric escape, and adopting the
  current mass, radius and orbital separation of \planet, we
  estimate mass loss rates of $2\--8\times10^{12}~g~s^{-1}$ during the
  saturated phase of X-ray emission (for efficiencies of
  5–25\%; \citealp{owen12,ionov18}). Assuming the X-ray evolution given by
  \citet{jackson12} and the corresponding extreme-ultraviolet
  emission by \citet{chadney15} and \citet{king18} we estimate a total mass
  loss of 2–9~\me. Considering instead the hydrodynamic calculations
  by \citet{kubyshkina18}, this mass loss increases to be greater than the total
  mass of the planet, and employing the detailed atmospheric escape evolution model of \citet{lopez17} suggests that the planet could have had an atmospheric mass fraction of up to $\sim$60\% of the total planet mass, or around half that of Saturn ($\sim$44~\me). This means that \planet\, could not have formed in situ with properties close to those we measure here, ruling out such a model. Conversely, adopting an initial planet mass and radius equal to that of Jupiter, we estimate a mass-loss of 5.5$\times10^{28}$~g over the current age of the system, which would only be $\sim$3\% of the total initial planet mass. Therefore, we can be sure that if the planet began as a Jupiter-mass gas giant, photoevaporation cannot be the sole mechanism that removed most of its atmosphere.

One possible mechanism for atmospheric loss is Roche Lobe Overflow
(RLO; \citealp{valsecchi15}). 
Planets with masses of $\sim$1~\mj\, orbiting solar-mass stars can fill their Roche Lobes for orbital periods approaching 12 hours.
For progenitor hot Jupiters with large cores ($\sim$30~\me), the
initial migration inwards to the RLO orbit is driven by tidal
interaction with the host star.  The migration can then reverse as mass
is stripped from the planet at a rate of 10$^{13} \-- 10^{14}$ g
s$^{-1}$ and continues on for a Gyr or so, assuming the escaping
material settles in an accretion disk around the star and transfers
its angular momentum back to the planet.  The planet can migrate outwards, reaching an orbital period of $\sim$0.8~days, before inward migration can resume.  Planets with
smaller masses undergo later inward migration within the mass loss
phase.  After the completion of RLO, these planets remain with an
atmosphere in the region of 7$\--$10\%, in agreement with that of
\planet, (assuming the planet is not still currently undergoing
  RLO).  Although these planets terminate with no atmosphere and an
orbital period of only 0.3~days after 2.1~Gyrs of evolution, commensurate
with the current age of \planet, less massive planets terminate with
orbital periods longer than more massive ones, and their mass loss
period increases also.  Although some of these models qualitatively fit the data observed for \planet, more work is still required to provide a stronger, more realistic description of the formation history of this system.  Finally, such a model is also dependent on the assumption that \planet\, started life as a gas giant planet, which is plausible
  given the planet's large heavy element abundance, and the fact that
  metal-rich stars are more commonly found to host gas giant planets
  than more metal-poor stars \citep{fischer05}.  

\newpage

%\small
\begin{longtable}{lcc}
\caption{\label{tab:star}Stellar properties of \starname}\\
  \hline
  Alternative Names \dotfill   &    \starname  &   \\
 &  	TIC\,183985250 & TESS \\
 &  	HIP\,117883 & HIPPARCOS \\
 & 2MASS J23544020-3737408 & 2MASS  \\
 & TYC 8015-1162-1 & TYCHO  \\
 \hline
Catalogue Data &&\\
RA \dotfill (J2000) &  23h54m40.60s &  TESS\\
DEC \dotfill (J2000) & -37d37m42.18s &   TESS\\
pm$^{\rm RA}$ \hfill (mas yr$^{-1}$) & 247.615 $\pm$ 0.076 & GAIA\\
pm$^{\rm DEC}$ \dotfill (mas yr$^{-1}$) & -69.801 $\pm$ 0.062 & GAIA\\
$\pi$ \dotfill (mas)& 12.403 $\pm$ 0.049 & GAIA \\ 
Photometric Data &&\\
T \dotfill (mag) & 9.10 $\pm$ 0.02 & TESS\\
B  \dotfill (mag) & 10.55 $\pm$ 0.04 & TYCHO\\
V  \dotfill (mag) & 9.76 $\pm$ 0.03 & TYCHO\\
G \dotfill (mag) & 9.6001 $\pm$ 0.0003 & GAIA\\
J  \dotfill (mag) & 8.45 $\pm$ 0.02 & 2MASS\\
H  \dotfill (mag) & 8.15 $\pm$ 0.02 & 2MASS\\
K$_s$  \dotfill (mag) & 8.02 $\pm$ 0.03 & 2MASS\\
WISE1  \dotfill (mag) & 7.94 $\pm$ 0.02 & WISE\\
WISE2  \dotfill (mag) & 8.02 $\pm$ 0.02 & WISE\\
WISE3  \dotfill (mag) & 8.00 $\pm$ 0.02 & WISE\\
\hline
Spectroscopic, Photometric, and Derived Properties &&\\
\teff  \dotfill (K) & 5445 $\pm$ 84 & \species\\
\logg \dotfill (dex) & 4.43 $\pm$ 0.31 & \species\\
\feh \dotfill (dex) & +0.25 $\pm$ 0.08 & \species\\
\vsini \dotfill (\kms) & 1.06 $\pm$ 0.37 & \species\\
\ensuremath{v_{mac}} \dotfill (\kms) & 1.98 $\pm$ 0.29 & \species\\
\teff  \dotfill (K) & 5496 $\pm$ 80 & \zaspe\\
\logg \dotfill (dex) & 4.51 $\pm$ 0.01 & \zaspe\\
\feh \dotfill (dex) & +0.24 $\pm$ 0.05 & \zaspe\\
\vsini \dotfill (\kms) & 1.7 $\pm$ 0.5 & \zaspe\\
\teff  \dotfill (K) & 5499 $\pm$ 50 & \spc\\
\logg \dotfill (dex) & 4.47 $\pm$ 0.10 & \spc\\
\meh \dotfill (dex) & +0.31 $\pm$ 0.08 & \spc\\
\vsini \dotfill (\kms) & 2.2 $\pm$ 0.5 & \spc\\
\teff  \dotfill (K) & \ensuremath{5443^{+14}_{-13}} & \ariadne\\
\logg \dotfill (dex) & \ensuremath{4.35^{+0.16}_{-0.12}} & \ariadne\\
\feh \dotfill (dex) & +0.27 $\pm$ 0.03 & \ariadne\\
\mstar \dotfill (\msun) & \ensuremath{1.03^{+0.03}_{-0.04}} & \species\, + MIST\\
\mstar \dotfill (\msun) & \ensuremath{1.00^{+0.02}_{-0.03}} & YY + GAIA\\
\mstar \dotfill (\msun) & \ensuremath{0.77^{+0.29}_{-0.21}} & \ariadne\\
\rstar \dotfill (\rsun) & 0.95 $\pm$0.01 & \species\, + MIST\\
\rstar \dotfill (\rsun) & 0.92 $\pm$ 0.01 & GAIA + This work\\
\rstar \dotfill (\rsun) & 0.949 $\pm$ 0.006 & \ariadne\\
L$_{\star}$ \dotfill (L$_{\odot}$) & \ensuremath{0.68 \pm 0.04} & YY + GAIA\\
L$_{\star}$ \dotfill (L$_{\odot}$) & \ensuremath{0.71 \pm 0.01} & \ariadne\\
M$_{V}$ \dotfill (mag) & 5.30 $\pm$ 0.07 & YY + GAIA\\
Age \dotfill (Gyr) & \ensuremath{2.1^{+2.2}_{-1.4}} & \species\, + MIST\\
Age \dotfill (Gyr) & \ensuremath{1.9^{+1.7}_{-1.2}} & YY + GAIA\\
$\rho_\star$ \dotfill (g cm$^{-3}$) & \ensuremath{1.81^{+0.06}_{-0.07}}  & YY + GAIA\\
Spectral Type   \dotfill  &  G7V   &  This work   \\
\ensuremath{<S_{HARPS}>} \dotfill & 0.148$\pm$0.008 & This work \\
\ensuremath{<\logrhk>} \dotfill & -5.10$\pm$0.04 & This work \\
\ensuremath{P_{rot,\vsini}} (days) \dotfill & 45 & This work \\
\hline
\end{longtable}

\newpage

   \begin{longtable}{lcc}
     %\left
   \caption{Transit, orbital, and physical parameters of \planet. For
  the prior descriptions, which are the expected probability
  distributions for each of the fit parameters, $N(\mu,\sigma)$
  represents a normal distribution with mean $\mu$ and standard
  deviation $\sigma$, whereas $U(a,b)$ and $J(a,b)$ represent a
  uniform prior and Jeffrey's prior, both defined between points $a$
  and $b$, respectively (see Methods: Global Modelling for more
  information).}\label{tab:planet} \\
  \hline
  Parameter      &   Prior &  Value \\
 \hline
 {\bf Light-curve parameters} & & \\
$P$ (days) & $N(0.792,0.1)$ & $0.7920520 \pm 0.0000093$\\
$T_0$ (days) & $N(2458354.22,0.1)$ & $2458354.21430 \pm 0.00025$\\
r$_{1}$ & $U(0,1)$ & \ensuremath{0.9417^{+0.0048}_{-0.0060}}\\
r$_{2}$ & $U(0,1)$ & \ensuremath{ 0.0454^{+0.0022}_{-0.0017}}\\
$\rho_{\star}$ (kg/m$^3$) & $N(1810,130)$ & \ensuremath{1758^{+125}_{-121}}\\
q$_{1,\tess}$ & $U(0,1)$ & \ensuremath{0.45^{+0.20}_{-0.16}}\\
q$_{2,\tess}$ & $U(0,1)$ & \ensuremath{0.43^{+0.35}_{-0.30}}\\
q$_{1,NGTS}$ & $U(0,1)$ & \ensuremath{0.63^{+0.25}_{-0.32}}\\
q$_{2,NGTS}$ & $U(0,1)$ & \ensuremath{0.55^{+0.31}_{-0.35}}\\
\hline
{\bf RV parameters} & & \\
K  (m s$^{-1}$)  &  $U(-100,100)$ & \ensuremath{19.65^{+0.43}_{-0.43}}\\
$e$              &  0    & 0 \\
$\omega$ (deg)   &  90  & 90 \\
$\gamma_{Coralie}$ (m s$^{-1}$) & $N(0,100)$ &   \ensuremath{-5.09^{+2.20}_{-2.20}} \\
$\gamma_{HARPS}$ (m s$^{-1}$)   & $N(0,100)$ &   \ensuremath{-4.40^{+0.30}_{-0.31}} \\
$\sigma_{Coralie}$ (m s$^{-1}$) & $J(10^{-2},100)$ &  \ensuremath{8.03^{+2.15}_{-1.74}} \\
$\sigma_{HARPS}$ (m s$^{-1}$)   & $J(10^{-2},100)$ &  \ensuremath{1.43^{+0.28}_{-0.24}} \\
\hline
{\bf Derived parameters} & & \\
$\rpl/\rstar$ & -- & \ensuremath{0.0455^{+0.0022}_{-0.0017}}\\
$a/\rstar$ & -- & \ensuremath{3.877^{+0.090}_{-0.091}}\\
$i$ & --  & \ensuremath{76.39 \pm 0.43} \\
\mpl\, (\me)     & -- & \mplan \\
\rpl\, (\re)      & -- & \rptess \\
$\langle$\teq\, $\rangle^{a}$ (K)         & -- & 1978 $\pm$ 19  \\
$a$  (AU)           &  -- & \ensuremath{0.01679^{+0.00014}_{-0.00012}}   \\
\rhopl\, (g cm$^{-3}$) &  -- & $1.536\pm0.123$   \\
\hline
  $^{a}$ Equilibrium temperature using equation 4 of \\
  \citet{mendez17} with $A=0.4$, $\beta=0.5$, and $\epsilon=1$.\\
\end{longtable}

\normalsize

Correspondence and requests for materials should be addressed to
James S. Jenkins (jjenkins@das.uchile.cl).

\subsection{Acknowledgements}
Funding for the TESS mission is provided by NASA's Science Mission directorate.
We acknowledge the use of public TESS Alert data from pipelines at the TESS Science Office and at the TESS Science Processing Operations Center.
This research has made use of the Exoplanet Follow-up Observation Program website, which is operated by the California Institute of Technology, under contract with the National Aeronautics and Space Administration under the Exoplanet Exploration Program.
Resources supporting this work were provided by the NASA High-End Computing (HEC) Program through the NASA Advanced Supercomputing (NAS) Division at Ames Research Center for the production of the SPOC data products.
JSJ and NT acknowledge support by FONDECYT grants 1161218, 1201371, and partial support from CONICYT project Basal AFB-170002. MRD is supported by CONICYT-PFCHA/Doctorado Nacional-21140646/Chile and Proyecto Basal AFB-170002.  JIV acknowledges support of CONICYT-PFCHA/Doctorado Nacional-21191829. 
This work was made possible thanks to ESO Projects 0102.C-0525 (PI: D\'iaz) and 0102.C-0451 (PI: Brahm).
RB acknowledges support from FONDECYT Post-doctoral Fellowship Project 3180246.
This work is partly supported by JSPS KAKENHI Grant Numbers JP18H01265 and JP18H05439, and JST PRESTO Grant Number JPMJPR1775.
The IRSF project is a collaboration between Nagoya University and the South African Astronomical Observatory (SAAO) supported by the Grants-in-Aid for Scientific Research on Priority Areas (A) (Nos. 10147207 and 10147214) and Optical \& Near-Infrared Astronomy Inter-University Cooperation Program, from the Ministry of Education, Culture, Sports, Science and Technology (MEXT) of Japan and the National Research Foundation (NRF) of South Africa.
We thank Akihiko Fukui, Nobuhiko Kusakabe, Kumiko Morihana, Tetsuya Nagata, Takahiro Nagayama, and the staff of SAAO for their kind support for IRSF SIRIUS observations and analyses.
CP acknowledges support from the Gruber Foundation Fellowship and Jeffrey L. Bishop Fellowship.
This research includes data collected under the NGTS project at the ESO La Silla Paranal Observatory. NGTS is funded by a consortium of institutes consisting of the University of Warwick, the University of Leicester, Queen's University Belfast, the University of Geneva, the Deutsches Zentrum f\" ur Luft- und Raumfahrt e.V. (DLR; under the `Gro\ss investition GI-NGTS’), the University of Cambridge, together with the UK Science and Technology Facilities Council (STFC; project reference ST/M001962/1 and ST/S002642/1). PJW, DB, BTG, SG, TL, DP and RGW are supported by STFC consolidated grant ST/P000495/1.
DJA gratefully acknowledges support from the STFC via an Ernest Rutherford Fellowship (ST/R00384X/1).
EG gratefully acknowledges support from the David and Claudia Harding Foundation in the form of a Winton Exoplanet Fellowship.
MJH acknowledges funding from the Northern Ireland Department for the Economy.
MT is supported by JSPS KAKENHI (18H05442, 15H02063).
AJ, RB, and PT acknowledge support from FONDECYT project 1171208,
and by the Ministry for the Economy, Development, and Tourism's
Programa Iniciativa Cient\'{i}fica Milenio through grant IC\,120009,
awarded to the Millennium Institute of Astrophysics (MAS).
PE, AC, and HR acknowledge the support of the DFG priority
program SPP 1992 “Exploring the Diversity of Extrasolar Planets” (RA
714/13-1).
We acknowledge the effort of Andrei Tokovinin in helping to perform the
observations and reduction of the SOAR data.

\subsection{Author Contributions}
JSJ led the \tess\, precision radial-velocity follow-up program, selection of the targets, analysis, project coordination, and wrote the bulk of the paper.  MD, NT, and RB performed the HARPS radial-velocity observations, PT observed the star with Coralie, and MD analysed the activity data from these sources.  NE performed the global modeling, with PCZ performing the TTV analysis, and RB, MGS, and AB performing the stellar characterisation using the spectra and evolutionary models.  PAPR worked on the \emperor\, code and assisted in fitting the HARPS radial-velocities.  EDL created a structure model for the planet, and in addition to  GWK and PJW, performed photoevaporation modeling.  JNW performed analysis of the system parameters.  DRC led the Keck NIRC2 observations and analysis. 
GR, RV, DWL, SS, and JMJ have been leading the \tess\, project, observations, organisation of the mission, processing of the data, organisation of the working groups, selection of the targets, and dissemination of the data products. 
CEH, SM, and TK worked on the SPOC data pipeline.
CJB was a member of the TOI discovery team.
SNQ contributed to TOI vetting, TFOP organization, and TRES spectral analysis. 
JL and CP helped with the interpretation of the system formation and evolution. 
KAC contributed to TOI vetting,  TFOP organization,  and TFOP SG1 ground-based time-series photometry analysis. GI, FM, AE, KIC, MM, NN, TN, and JPL contributed TFOP-SG1 observations.
JSA, DJA, DB, FB, CB, EMB, MRB, JC, SLC, AC, BFC, PE, AE, EF, BTG, SG, EG, MNG, MRG, MJH, JAGJ, TL, JM, MM, LDN, DP, DQ, HR, LR, AMSS, RHT, RTW, OT, SU, JIV, SRW, CAW, RGW, PJW, and GWK are part of the NGTS consortium who provided follow-up observations to confirm the planet.  
EP and JJL helped with the interpretation of the result.  
CB performed the observations at SOAR and reduced the data, CZ performed the data analysis, and NL and AWM assisted in the survey proposal, analysis, and telescope time acquisition.
All authors contributed to the paper.

\subsection{Author Information}
Reprints and permissions information is available at \texttt{www.nature.com/reprints}. Correspondence and requests for materials should be addressed to James S. Jenkins, \texttt{jjenkins@das.uchile.cl}

\subsection{Competing Interests}
The authors declare that they do not have any competing financial interests.

\clearpage
\pagebreak

\section{Extended Data}

\begin{figure}
\center
\includegraphics[width=\textwidth]{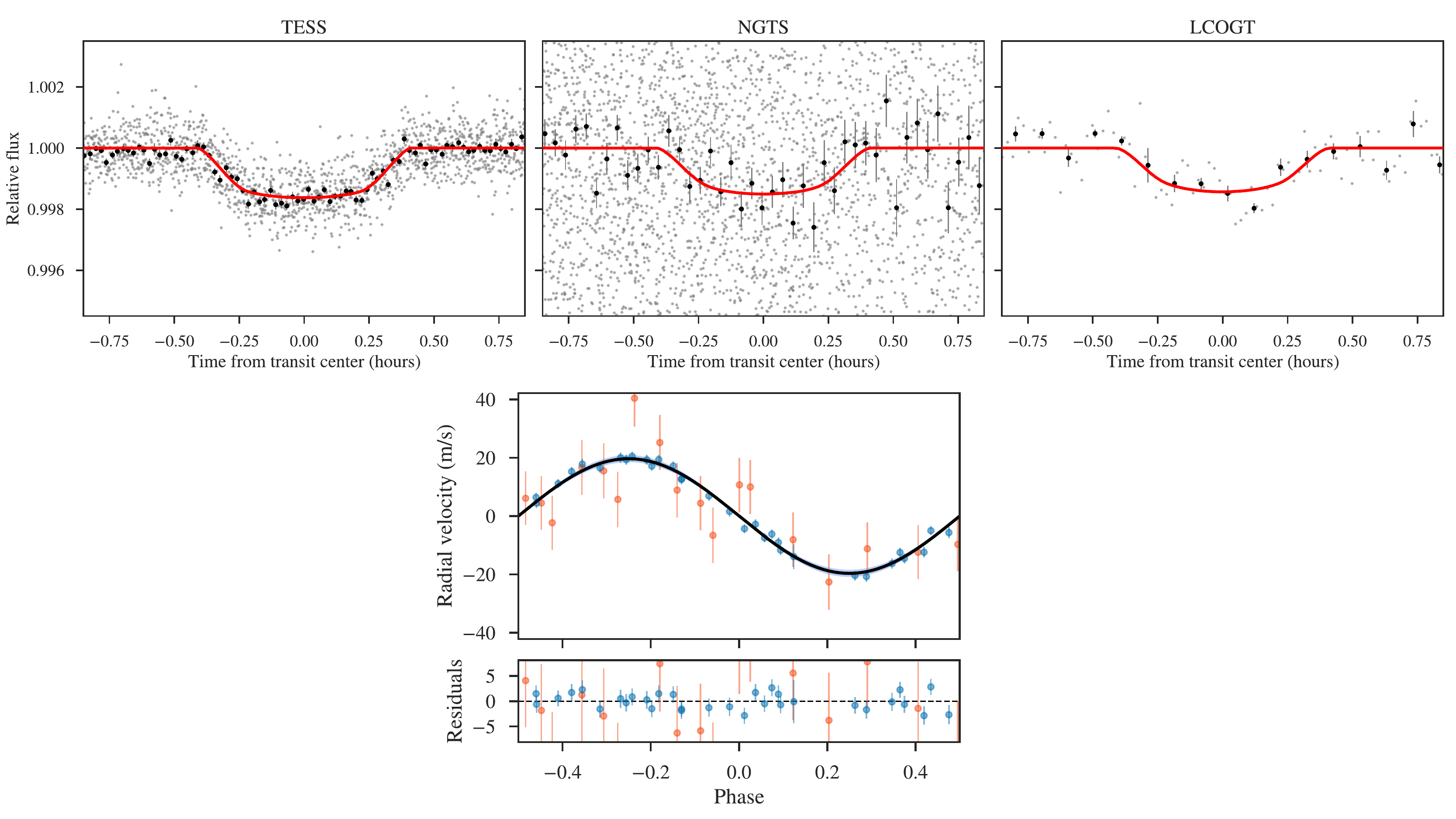} %\label{fig:lc}
\caption{{\bf Transit lightcurves and phase folded RVs for \starname} The left top panel shows the discovery \tess\ lightcurve averaging over 33 transits, with the center and right panels showing the single transit follow-up photometry from NGTS and LCOGT for \starname, with a 0.8~day period, and including the associated transit model using the parameters shown in Table~\ref{tab:planet}.  The full data set is shown by the small points and the binned data is superimposed on these as the larger and darker points with associated uncertainties. The lower panel shows the 31 HARPS radial-velocities in blue and 18 Coralie measurements in orange (see Table~\ref{tab:rvs}), also folded to the period of the planet, and with their respective uncertainties.  The mean uncertainty of the Coralie velocities is a factor of 10 larger than those from HARPS.  The best fit model is again overplotted on the data.  The \tess\, NGTS and LCOGT photometry, along with the HARPS and Coralie velocities were fit simultaneously to ensure the best constraints possible on the planet parameters, along with a more accurate description of the overall uncertainties.
}
\end{figure}

\begin{figure}
\includegraphics[width=1.0\textwidth]{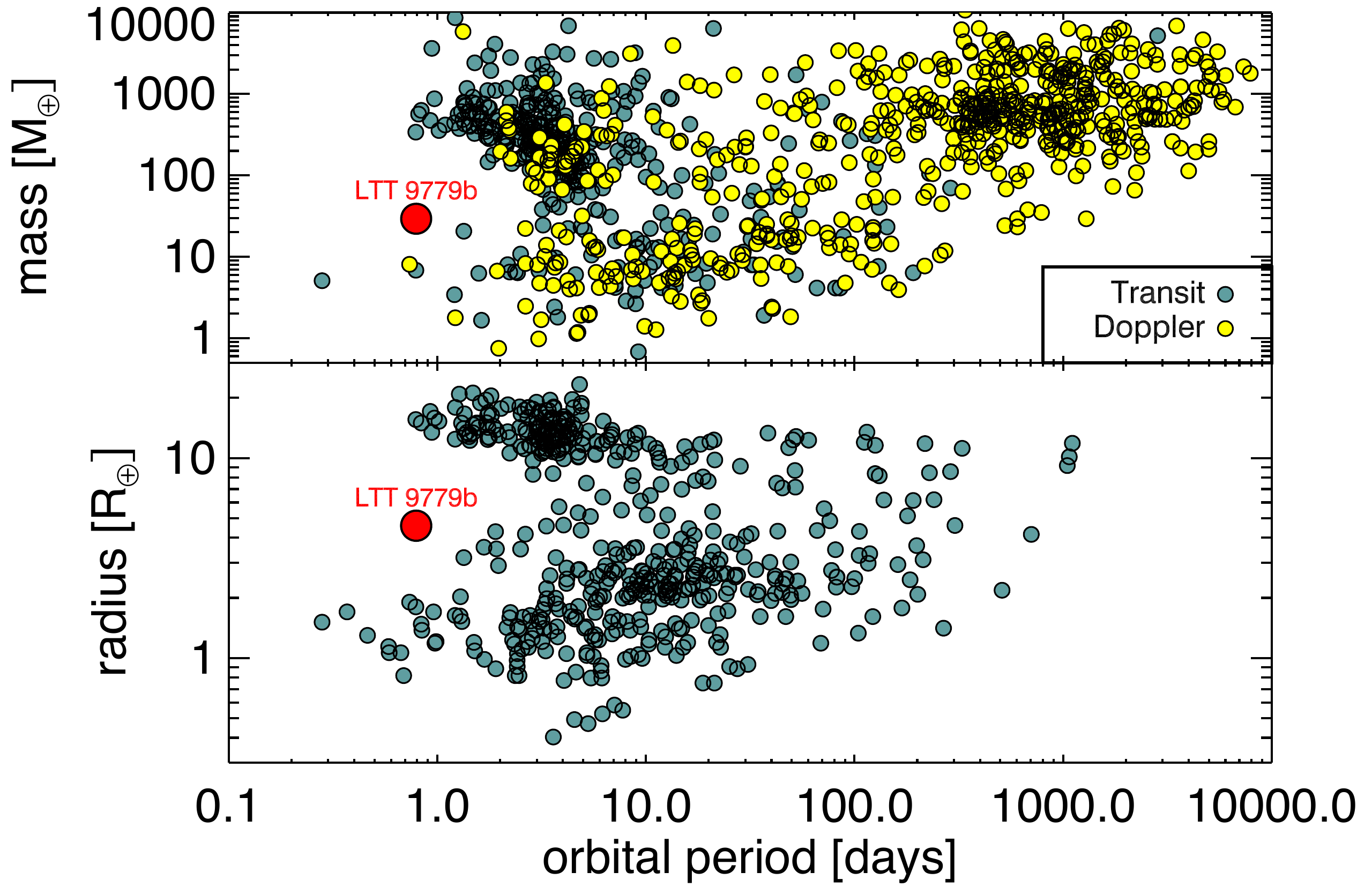} %\label{fig:per_mass}
\caption{{\bf \planet\, in the period\--mass and period\--radius planes.} The top plot shows all currently confirmed planets with a fractional mass uncertainty below 30\%, separated in colour by their detection method.  The lower plot shows all currently confirmed transit planets with a fractional radius uncertainty below 5\%.  \planet\ is clearly isolated in the Neptune desert in period\--mass\--radius space, meaning it is heavily decoupled from the current populations of known exoplanets.
}

\end{figure}

\begin{figure}
\includegraphics[width=1.0\textwidth]{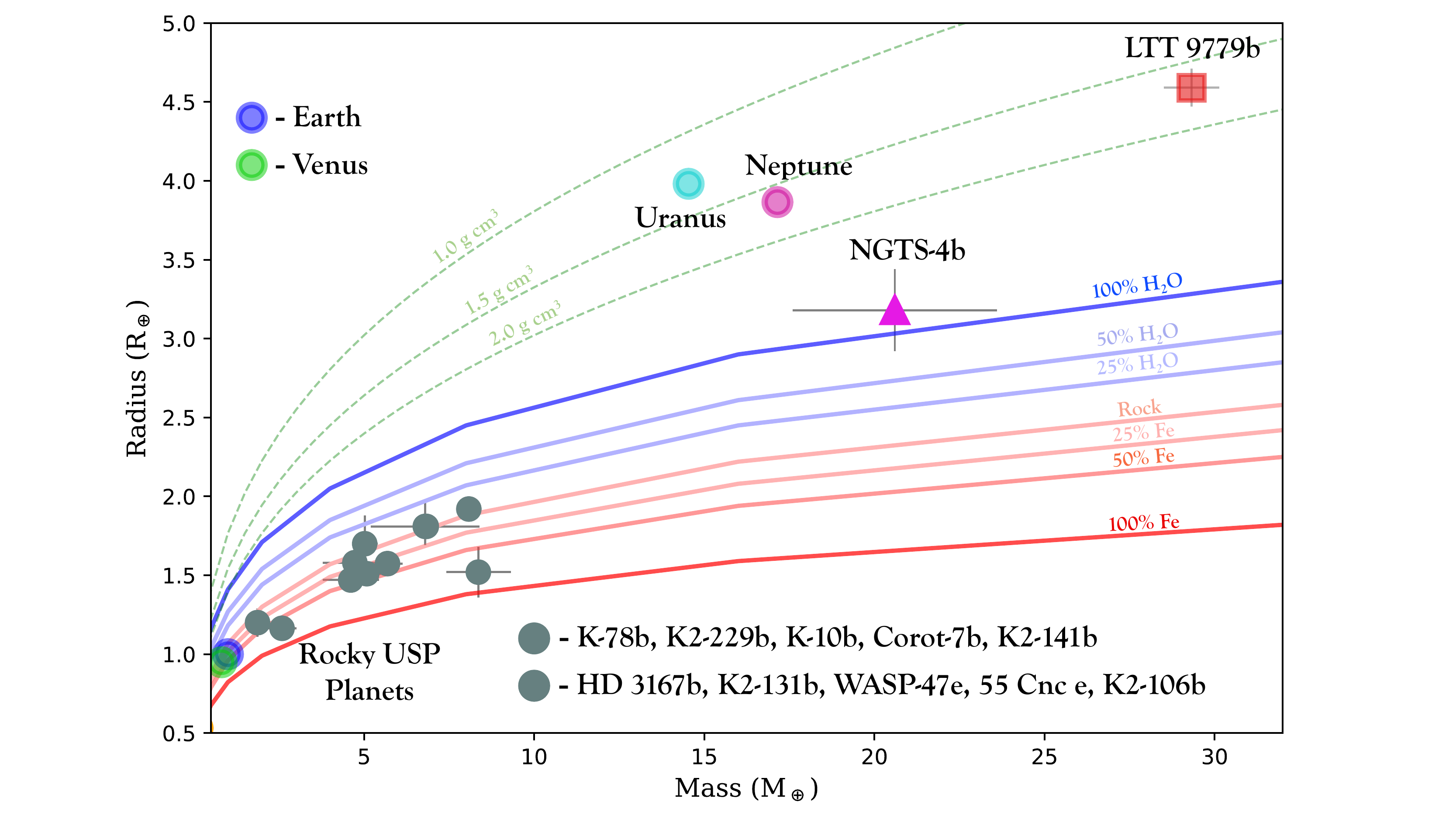} %\label{fig:mass_rad}
\caption{{\bf \planet\, in the mass\--radius plane.}  The plot includes all non-gas giant ultra-short period planets with well constrained Doppler masses.  \planet\ is marked by the red square.  Structure models from \citet{Zengetal2016} are plotted as solid curves and labelled depending on the bulk composition of the planet.  The models range from a 100\% iron core planet, through to a 100\% water world.  The ultra-short period planets all agree with rocky-iron compositions, explained by photoevaporation of their primordial atmospheres.  \planet\ is significantly larger, indicating that it has a residual hydrogen and helium atmosphere around the core.  Dashed iso-density curves are shown in green for reference, highlighting the similar densities between Neptune and \planet.  For reference, Venus, Earth, Uranus, and Neptune are represented in the plot.
}

\end{figure}

\tiny
\begin{figure}
\includegraphics[width=1.0\textwidth]{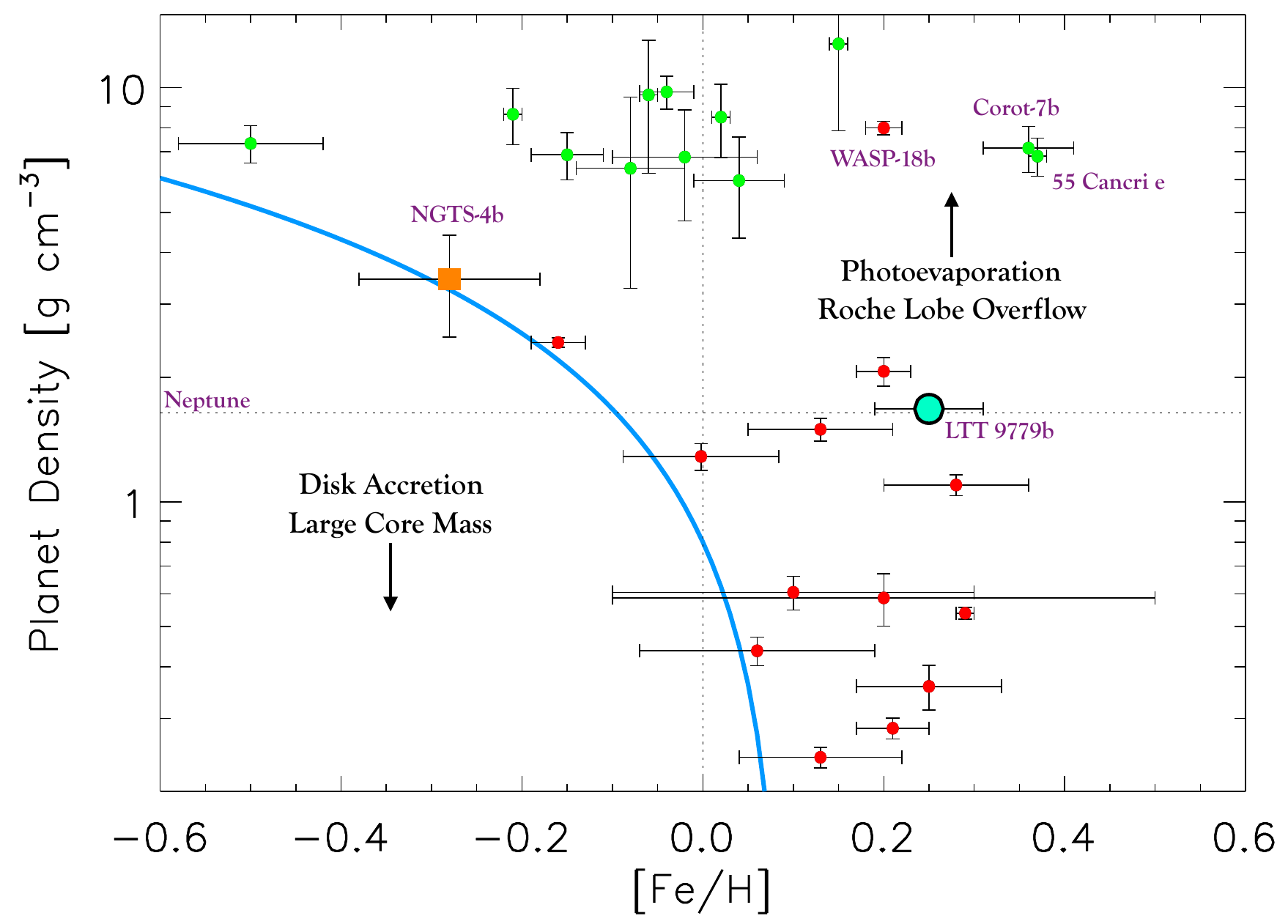} %\label{fig:mass_rad}
\caption{{\bf Distribution of planetary densities as a function of host star metallicity for currently known transiting planets with orbital periods less than 1.3 days.} The sample is split into those with masses less than 0.1\mj\, (green circles; ultra-short period planet proxies) and those with masses above (red circles; ultra-hot Jupiter proxies).  The ultra-hot Neptune \planet\, and longer period NGTS-4~b are clearly labelled in the figure.  The blue curve is a power law described by $3.5 \times [(N_{Fe}/N_{H})/(N_{Fe}/N_{H})_{\odot}]^{-2.5} + 0.8$, which bounds the regions governed by the physical processes that determine the planet bulk properties. Those physical processes and the direction in which they move planets are shown in the plot.
}
\end{figure}

\normalsize

\clearpage
\pagebreak

\section{Methods}

\noindent{\bf \tess\, Photometry Treatment}

\tess\, observed the star \starname\, (HIP\,117883, TIC\,183985250,
TOI\,193) using Camera 2 (CCD 4), between UT 2018 Aug 23 and Sep 20
(JD 2458354.11439 $\--$ 2458381.51846), part of the Sector 2 observing
campaign.  The short cadence time sampling of the data was set to two
minutes, and data products were then processed on the ground using the
Science Processing Operations Center (SPOC) pipeline package
\citep{jenkins16_spie}, a modified version of the Kepler mission
pipeline \citep{smith12,stumpe14,twicken18,li19}.  SPOC delivers Data Validation Reports to MIT.  These reports document so-called "Threshold Crossing Events" identified by the SPOC pipeline, namely dips that could conceivably be due to transiting planets.  A team in the TESS Science Office then reviews the reports, including analogous reports from analysis of the Full Frame Images using the Quick-Look Pipeline developed at MIT.  There are many criteria that go into deciding whether a TCE is an instrumental artifact or false alarm (e.g. due to low SNR), or an astrophysical false positive due to eclipsing binaries contaminating the TESS photometry or otherwise masquerading as transiting planets.  This is the so-called vetting process.  Candidates that survive vetting are then assigned TOI numbers and announced to the public, e.g. at MAST, and then work by the TESS Follow-up Observing Program Working Group begins, tracking down TOIs that are not planets but could not be rejected based on the information available to the vetting team.  TOIs that survive the reconnaissance work of the TFOP WG can then move on to precision RV work.  \planet\, was released as a \tess\, alert on the 4th of October 2018, and assigned the code \tess\, Object of Interest (TOI) 193.  As part of the alert process candidate vetting, the light curve modeling did not show any hints of abnormality, such that no transit depth variations were apparent, no PSF centroiding offsets were found, and no secondary eclipses report, giving rise to a bootstrap false alarm probability (FAP) of 2$\times$10$^{-278}$.

\begin{figure}
\includegraphics[width=1.0\textwidth]{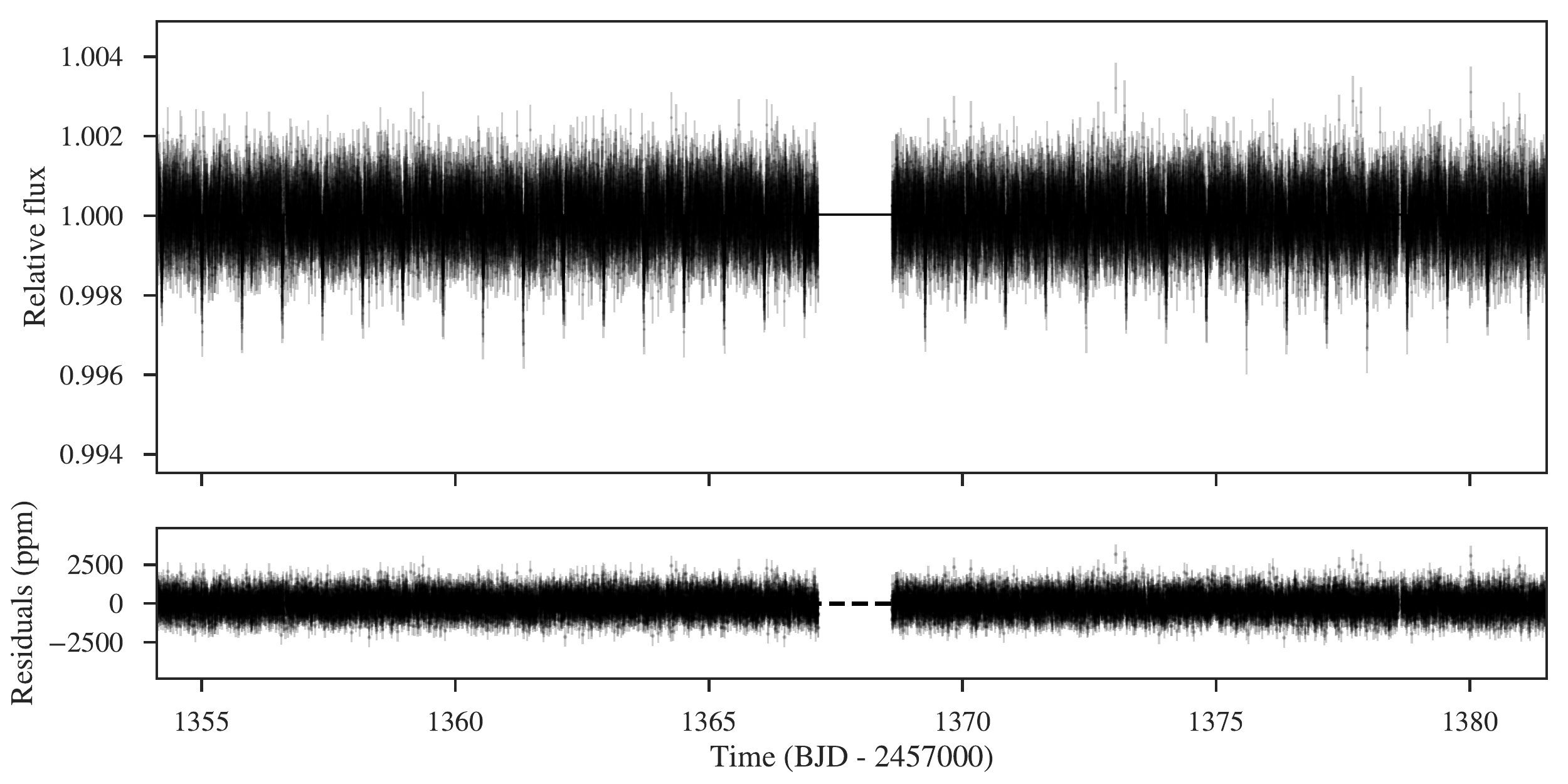}\label{fig:tesslc}
\caption{{\bf Normalised \tess\, pre-search data conditioning timeseries photometry for \starname\,} with the optimal model (black curve) overplotted on the data (top).  The model residuals are shown in the lower panel.
}
\end{figure}

In Figure~5 we show the \tess\, pre-search data conditioning light curve for \starname, after removal of points that were flagged as being affected by excess noise.  Given the quiescent nature of the star, the photometric light curve is fairly flat across the full time series, with the small transits (\tessdepth) readily apparent to the eye.  This simplified the modeling effort, giving rise to the small residual scatter shown in the figure.
\\

\noindent{\bf Follow-up NGTS Photometry}

Photometric follow-up observations of a full transit of \planet\ was
obtained on UT 2018 Dec 25 using the Next Generation Transit Survey
(NGTS) at ESO's Paranal observatory \citep{wheatley18}.  We used a new
mode of operation in which nine of the twelve individual NGTS
telescopes were used to simultaneously monitor \starname\
\citep{smith20}.  We find that the photometric noise is uncorrelated
between the nine telescopes, and therefore we improve the photometric
precision by a factor of three compared to a single NGTS telescope.
The observations were obtained in photometric conditions and at
airmass $< 2$.  A total of 6502 images were obtained, each with an
exposure time of 10\,s using the custom NGTS filter
(520$\--$890\,nm). The observations were taken with the telescope
slightly defocussed to avoid saturation. The telescope guiding was
performed using the DONUTS auto-guiding algorithm \citep{McCormac13}, which resulted in an RMS of the target location on the CCD of only 0.040\,pix, or 0.2\,$''$,. Due to this high precision of the auto-guiding, the use of flat fields during the reduction of the images was not required. Comparison stars were chosen manually and aperture photometry was performed on the images using a custom aperture photometry pipeline. The wide field-of-view provided by NGTS enabled the selection of a good number of suitable comparison stars, despite \starname\ being a relatively bright star. When combined, the resulting photometry showed the transit signal of TOI-193 with a depth and transit centre time consistent with the \tess\ photometry. The combined NGTS light curve has a precision of 170\,ppm over a half hour timescale, which is a comparable to the \tess\, precision of 160\,ppm over this timescale (for a single transit).
\\

\noindent{\bf Dilution Probability}

Given the reality of the transit as 'on-source', the issue of dilution of the light curve by a foreground or background star is considered in a probabilistic sense.  In this case, we aim to test the probability of having a blended star so close to the star angularly on the sky, that the AO observations would not have detected it.  The AO sensitivity deteriorates quickly below 0.5$''$ or so, with low sensitivity to objects with angular separations of 0.1$''$ or less on sky.

\subsection{Background or Foreground Contaminant}

With this in mind, we used the Besan\c{c}on galactic model \citep{robin03} to generate a representative star field around the position of \starname, with the aim of testing the likelihood of having a diluted star that significantly affects the transit parameters.  The model has been used in a similar manner previously.  For instance, in \citet{fressin13} they applied the model to test the probability that each of the Kepler transit planet candidates in their study was the result of a blended eclipsing binary.  We selected all stars within a one square degree box surrounding our target, down to the magnitude limit of $V = 21$ that the model provides.  This gave rise to over 2200 stars to work with, for which we randomly assign positions in RA and Dec using a uniform random number generator, constrained to be within the selected box boundaries.  We then ran the simulation 10 million times to generate a representative sample, recording all the events where a star passed within a separation of 0.1$''$ from \starname, and finally normalising by the number of samples probed.  The test returned only 48 events, providing a probability to have such a close separation between two stars in this field of only 4.8 $\times$ 10$^{-6}$ (0.0005\%).  

Although the probability we found is very small, it is actually an upper limit.  Blending by stars as faint as 21st magnitude for instance, does not affect the transit depth enough to push the radius of the planet above the Neptune desert.  The faintest population of stars in our test, which also represents the most abundant population, biases the probability to larger values.  For instance, if we take the mean of the final bin (20.5 magnitudes), we have a magnitude difference from \starname\, of 10.2, which relates to an effect at the level of 83~ppm, only 5\% of the observed transit depth.  If \planet\, is truly a hot Jupiter, then in order to push it out of the Neptune desert, given its orbital period and current radius, we require dilution from a star of $\sim$5.5 magnitudes or brighter, limiting our test to only stars in magnitude bins of 16 or less.  Performing this test decreases further the probability down to 1 $\times$ 10$^{-6}$ (0.00001\%), ruling out the possibility that dilution of the light curves is the reason the planet falls in such an isolated part of the parameter space.

\subsection{Binary Star Contaminant}

Although a non-bound stellar contaminant is unlikely to be diluting the transit of \planet\, sufficiently to push it out of the Neptune desert, a binary companion is likely to have a higher chance to be present and tightly separated to \starname.  Therefore, we performed Monte Carlo simulations to test how likely having a stellar binary that is bright enough to dilute the transit lightcurve sufficiently would be.  We simulated 10$^5$ binary systems, drawing the system parameters from the probability density functions (PDFs) calculated in \citet{raghavan10}.  Here the orbital log-period PDF in days is a normal distribution with mean of 5.03 and standard deviation of 2.28.  Most other parameters like eccentricity, the orbital angles, and the mass ratio, were simulated using uniform distributions within their respective bounds.  Only the system inclinations were drawn from a cosine PDF.  

When simulating the systems, we normalized each by the fraction of the orbital period that the secondary star would spend within 0.1$''$ of the primary.  Therefore, systems that never approached within this angular separation were assigned a fractional time ($T_{f}$) of zero, those that always were found within this limit were assigned a value of unity, and the rest were assigned a value between 0$\--$1 depending on the fractional time spent within this distance.  With these calculations we could apply the formulism \ensuremath{P = (\sum_{n=1}^{n=N_t} P_{n}T_{f,n}) / N_t}, where the probability $P$ is the sum total of fractional probabilities $P_n T_{f,n}$ across all samples, normalised by the total number of samples $N_t$.  Finally, we then normalised by the 46\% fraction of such stars found to exist in binaries.  

With these simulations we arrived at a value of 13.5\% for the probability that \starname\, has a binary companion that could be found within an angular separation of 0.1$''$ at any one time.  Although this is a relatively large probability, this is integrated across all binary mass fractions, and therefore does not take into account that only a small mass range is permitted by the spectral analysis.  When we account for the cross correlation function analysis discussed below, the probability drops to essentially zero, since the larger secondary masses required to affect the transit depth sufficiently are all ruled out.
\\

\noindent{\bf Gaia Variability}

Another way to probe for very closely separated stars on the sky is to
study the measurements made by Gaia, in particular the excess noise
parameter $\epsilon$ and the Tycho-Gaia astrometric solution (TGAS)
discrepancy factor $\Delta Q$ \citep{lindegren12,michalik14,lindegren16}.  These can be used to look for excess variability in the observations that are indicative of blended starlight from a foreground or background star, spatially close enough that they can not be resolved by the instrument.

Both the $\epsilon$ and $\Delta Q$ parameters are listed in Gaia DR1
as standard outputs, however Gaia DR2 only reports the excess noise,
which turns out to be unreliable for stars with G $<\sim$13
\citep{lindegren18}. $\Delta Q$ measures the difference between the
proper motion derived in TGAS and the proper motion derived in the
Hipparcos Catalog \citep{leeuwen07}. Also, \citet{rey17} utilised both $\Delta Q$ and $\epsilon$ to show the lack of binarity for some stars. $\Delta Q$ is expected to follow a $\chi^2$ distribution with two degrees of freedom for single stars. The Gaia DR1 $\epsilon$ and $\Delta Q$ for LTT9779 are 0.394 (with a significance of 134.281) and 2.062, respectively. According to \citet{lindegren16}, all sources obtain a significant excess source noise of $\sim$0.5 mas, due to poor attitude modeling (so an excess noise $>$ 1 - 2 could indicate binarity), and a significance $>$ 2 indicates that the reported excess noise is significant, therefore, from excess noise alone, \starname\, is astrometrically well behaved and shows no evidence of binarity. While \citet{michalik14} reports a $\Delta Q$ threshold of 15.086 for a star to be well behaved (at a significance level of 1\%), \citeauthor{lindegren16} reduces this threshold to 10. This means that any star with $\Delta Q$ $<$ 10 is considered to be astrometrically well behaved, again showing that this star is highly likely to be uninfluenced by contaminating light from a background object.
\\

\noindent{\bf Follow-up Spectroscopy}

\subsection{NRES Spectroscopy}

In order to aid in characterisation of the host star we used the LCO
robotic network of telescopes \citep{brown13} and the Robotic Echelle
Spectrographs (NRES; \citealp{siverd18}). We obtained 3 spectra, each composed of 3 x 1200 sec exposures, on UT 2018 Nov 5, 8, and 9. All three spectra were obtained with the LCO/NRES instrument mounted on a 1~m telescope at the LCO CTIO node. The data were reduced using the LCO pipeline resulting in spectra with SNR of 61$\--$73. We have analyzed the spectra using SpecMatch while incorporating the Gaia DR2 parallax using the method described by \citet{fulton18}. The resulting host stars parameters contributed to those listed in Table~\ref{tab:star}.

\subsection{TRES Spectroscopy}

We obtained two reconnaissance spectra on the nights of UT2018-11-04
and UT2018-11-05 using the Tillinghast Reflector Echelle Spectrograph
(TRES; \citealp{furesz08}) located at the Fred Lawrence Whipple
Observatory (FLWO) in Arizona, USA. TRES has a resolving power of
$\sim$44,000, covering a wavelength range of 3900$\--$9100\AA, and the
resulting spectra were obtained with SNRs of $\sim$35 at 5200\AA.  The
spectra were then reduced and extracted as described in
\citet{buchhave10}, whereby the standard processing for echelle
spectra of bias subtraction, cosmic ray removal, order-tracing,
flatfielding, optimal extraction \citep{horne86}, blaze removal,
scattered-light subtraction, and wavelength calibration was applied.
The Spectral Parameter Classification tool \citep{buchhave12} was used to measure the stellar quantities we show in Table~\ref{tab:star}.

\subsection{HARPS Spectroscopy}

Upon examination of the light curve we decided to perform high cadence
follow-up spectroscopic observations with the High-Accuracy Radial
velocity Planet Search spectrograph (HARPS; \citealp{pepe00}) installed at the ESO 3.6m telescope in La Silla, in order to fully cover the phase space.  We started observing LTT 9779 on Nov 6th 2018. From an initial visual examination of the spectra and cross correlation function (CCF) of the online Data Reduction Software (DRS), it was consistent with no evidence of blending with other stellar sources nor as being a fast rotator or active, based on the width of the CCF and various activity indicators.

We acquired 32 high-resolution (R$\sim$ 115,000) spectroscopic
observations between 2018 Nov 6 and Nov 9, Dec 11 to 13, and Dec 28 to
Dec 30, where for the nights of Nov 7,8 and 9th we observed the star
four times throughout the night to fully sample the orbital period. We
integrated for 1200s, using a simultaneous Thorium-Argon lamp
comparison source feeding fiber B, and we achieved a mean
signal-to-noise ratio of $\sim$65.4.  We then reprocessed the
observations using the HARPS-TERRA analysis software \citep{AngladaEscudeEtal2012} where a high-signal to noise template is constructed from all the observed spectra. Then the radial velocities are computed by matching each individual observation to the template, and we list these measurements in Table~\ref{tab:rvs}. 

HARPS-TERRA receives the observed spectra, stellar coordinates, proper
motions, and parallaxes as input parameters. The output it produces
then consists of a series of radial velocities that are calculated for
a given wavelength range in the echelle. We used the weighted
radial-velocities that were calculated starting from the 25th order of
the HARPS echellogram, which is centered at a wavelength of 4500 \AA.
We chose this wavelength range since the uncertainties it produced had
the lowest MAD\footnote{Median Absolute Deviation = median(\ensuremath{|x_{i}-median(x)|}} value.  This is likely due to increased
stellar activity noise that affects the bluest orders the most,
combined with relatively low signal-to-noise ratios, and therefore
removing these orders allows higher precision to be reached. It was
with this data that we performed the \emperor\ \citep{JenkinsEtal2018}
fitting, providing the independently confirmed and constrained
evidence for \planet\, (Figure~6).  For instance, the Doppler orbital
period was found to be 0.7920 $\pm 0.0001~d$, in excellent agreement
with that provided by the TESS transit fitting, and allowing the
period to be constrained in the joint fit to one part in 80'000
(0.001\%).  We also used this spectra to test if possible spectral
line asymmetries and/or activity related features could be driving the
signal.  In particular, we searched for linear correlations between
the spectral bisector inverse slope measurements and the
radial-velocities (see Figure~7), along with performing period
searches using Generalized Lomb Scargle periodograms \citep{zechmeister09} and Bayesian methods with the \emperor\, code.  The Spearman correlation coefficient between the BIS and RVs is found to be 0.22 with a p-value of 0.22, meaning there exists no strong statistical evidence to reject the null hypothesis that such a weak  correlation has arisen by chance.  From the periodogram analyses, no statistically significant periodicities were detected with false alarm probabilities of less than 0.1\%, our threshold for signal detection. We also performed the same analyses on the full width at half maximum of the HARPS cross-correlation function, and chromospheric activity indicators like the $S$, H$\alpha$, and HeI indices, again with no statistically significant results encountered.  

Finally, we also reprocessed the HARPS spectra to generate CCFs with binary masks optimised for spectral types between G2-M4, but across a wider $\pm$200~\kms range in velocity to check for weaker secondary CCFs that could be due to additional, nearby companions.  We took a typical HARPS \starname\, spectrum and injected mid-to-late M star spectra with decreasing SNRs, until we could not detect the M star CCFs no more, providing an upper limit on the mass of any contaminating secondary.  From analysis of the mean flux ratio between the M stars and \starname, we found that we should be able to detect stellar contaminants down to a mass of 0.19~\msun, using the mass-luminosity relation of \citet{benedict16}, however no companion CCFs were detected.  Such a companion would have a magnitude difference of over 7.5, and since we previously calculated above that a maximum magnitude difference of 5.5 would be required to push LTT9779b out of the Neptune Desert, the limits permitted by the CCF analysis show that a diluted companion would not change the conclusions of our work.

\subsection{Coralie Spectroscopy}

Additional phase coverage was performed using the Coralie spectrograph installed in the 1.2~m Swiss Leonhard Euler Telescope at the ESO La Silla Observatory in Chile.  Coralie has a spectral resolution of $\approx$60000 and uses a simultaneous calibration fibre illuminated by a Fabry-Perot etalon for correcting the instrumental radial-velocity drift that occurs during the science exposures.
The star was observed a total of 18 times throughout the nights of
2018 Nov 15 to Nov 20. The adopted exposure time for the Coralie
observations was 1200s, and the SNR obtained per resolution element at
5150 \AA\, ranged between 50 and 60. Coralie data was processed with
the CERES pipeline \citep{brahm17}, which performs the optimal extraction of
the science and calibration fibres, the wavelength calibration and instrumental drift correction, along with the measurement of precision radial-velocities and bisector spans by using the cross-correlation technique. Specifically, a binary mask optimised for a G2-type star was used to compute the velocities for \starname. The typical velocity precision achieved was $\approx$5~ms$^{-1}$, which allowed the identification the Keplerian signal with an amplitude of 20~$^{-1}$.\\

\small
\begin{longtable}{cccc}
\caption{\label{tab:rvs}Radial-velocities of \starname}\\
\hline
JD - 2450000 & RV & Uncertainty & Instrument \\
 & (m s$^{-1}$) & (m s$^{-1}$) & \\
 \hline
8429.51804 & -10.59 & 0.86 & HARPS \\
8430.54022 & -16.91 & 0.74 & HARPS \\
8430.59553 & -9.41 & 0.68 & HARPS \\
8430.67911 & 1.99 & 0.79 & HARPS \\
8430.76201 & 13.40 & 1.21 & HARPS \\
8431.51068 & 6.71 & 0.61 & HARPS \\
8431.64346 & 16.09 & 0.83 & HARPS \\
8431.69130 & 14.98 & 0.87 & HARPS \\
8431.73217 & 8.41 & 0.55 & HARPS \\
8432.50941 & 12.77 & 0.73 & HARPS \\
8432.65689 & -7.23 & 0.94 & HARPS \\
8432.69804 & -13.45 & 1.06 & HARPS \\
8432.72573 & -18.32 & 4.02 & HARPS \\
8464.53817 & -25.17 & 1.02 & HARPS \\
8464.64153 & -16.81 & 1.11 & HARPS \\
8464.68616 & -10.08 & 1.27 & HARPS \\
8465.53024 & 0.00 & 0.85 & HARPS \\
8465.59314 & 10.82 & 0.84 & HARPS \\
8465.64411 & 12.09 & 0.86 & HARPS \\
8465.68104 & 15.61 & 1.12 & HARPS \\
8466.52022 & 14.89 & 1.03 & HARPS \\
8466.58232 & 8.12 & 0.90 & HARPS \\
8466.63157 & 2.49 & 1.09 & HARPS \\
8466.66865 & -2.85 & 1.10 & HARPS \\
8481.53213 & 14.93 & 0.94 & HARPS \\
8481.57805 & 12.72 & 0.84 & HARPS \\
8482.53643 & -8.75 & 0.74 & HARPS \\
8482.57255 & -11.89 & 0.82 & HARPS \\
8482.60140 & -16.09 & 0.90 & HARPS \\
8483.52686 & -24.82 & 0.80 & HARPS \\
8483.59338 & -20.68 & 1.12 & HARPS \\
8483.61557 & -18.95 & 0.93 & HARPS \\
\hline
8438.56440 & -14.80 & 4.50 & CORALIE  \\
8438.62857 & -7.40  & 4.60  & CORALIE \\
8438.72084 & 10.40 & 5.00 & CORALIE \\
8439.56828 & 35.30 & 5.60 & CORALIE \\
8439.64481 & 3.80 & 4.80 & CORALIE \\
8439.70910 & -11.70 & 5.20 & CORALIE \\
8440.56824 & 4.90 & 4.70 & CORALIE \\
8440.64498 & -13.20 & 4.70 & CORALIE \\
8440.70927 & -27.70 & 5.00 & CORALIE \\
8441.57027 & -16.30 & 4.20 & CORALIE \\
8441.66132 & -17.50 & 4.60 & CORALIE \\
8441.74898 & 1.00 & 4.50 & CORALIE \\
8442.56932 & -0.60 & 4.50 & CORALIE \\
8442.64202 & 11.60 & 4.90 & CORALIE \\
8442.70651 & 0.60 & 5.00 & CORALIE \\
8443.57400 & 20.10 & 5.00 & CORALIE \\
8443.64711 & -0.70 & 4.70 & CORALIE \\
8443.71686 & 5.60 & 4.80 & CORALIE \\
\hline
\end{longtable}
\normalsize

\begin{figure}
\center
\includegraphics[width=0.6\textwidth,angle=270]{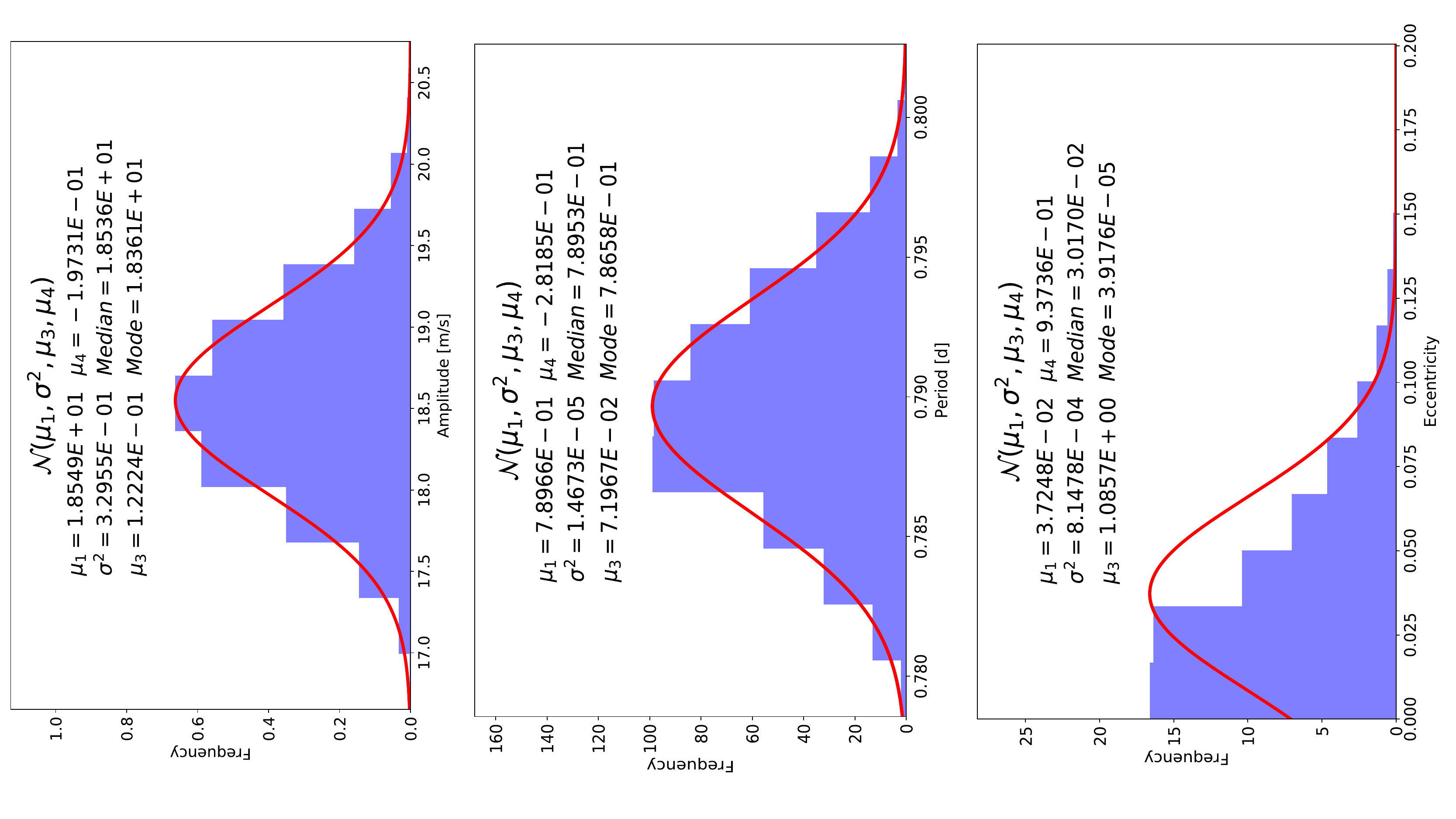}\label{fig:rvs}
\caption{{\bf Independently constrained system parameters from the \emperor\, MCMC runs of the 31 HARPS radial-velocities.}  From top to bottom we show the posteriors of the velocity amplitude, the orbital period, and the eccentricity of the orbit.  Overplotted on each histogram is a gaussian distribution with the same input parameters as those calculated from the posterior distributions.  We also show the values obtained from the distributions.  The histograms reveal that the signal is well constrained with the current data in hand, and the period in particular is in excellent agreement with that from the \tess\ lightcurve.
}
\end{figure}

\begin{figure}
\center
\includegraphics[width=0.6\textwidth]{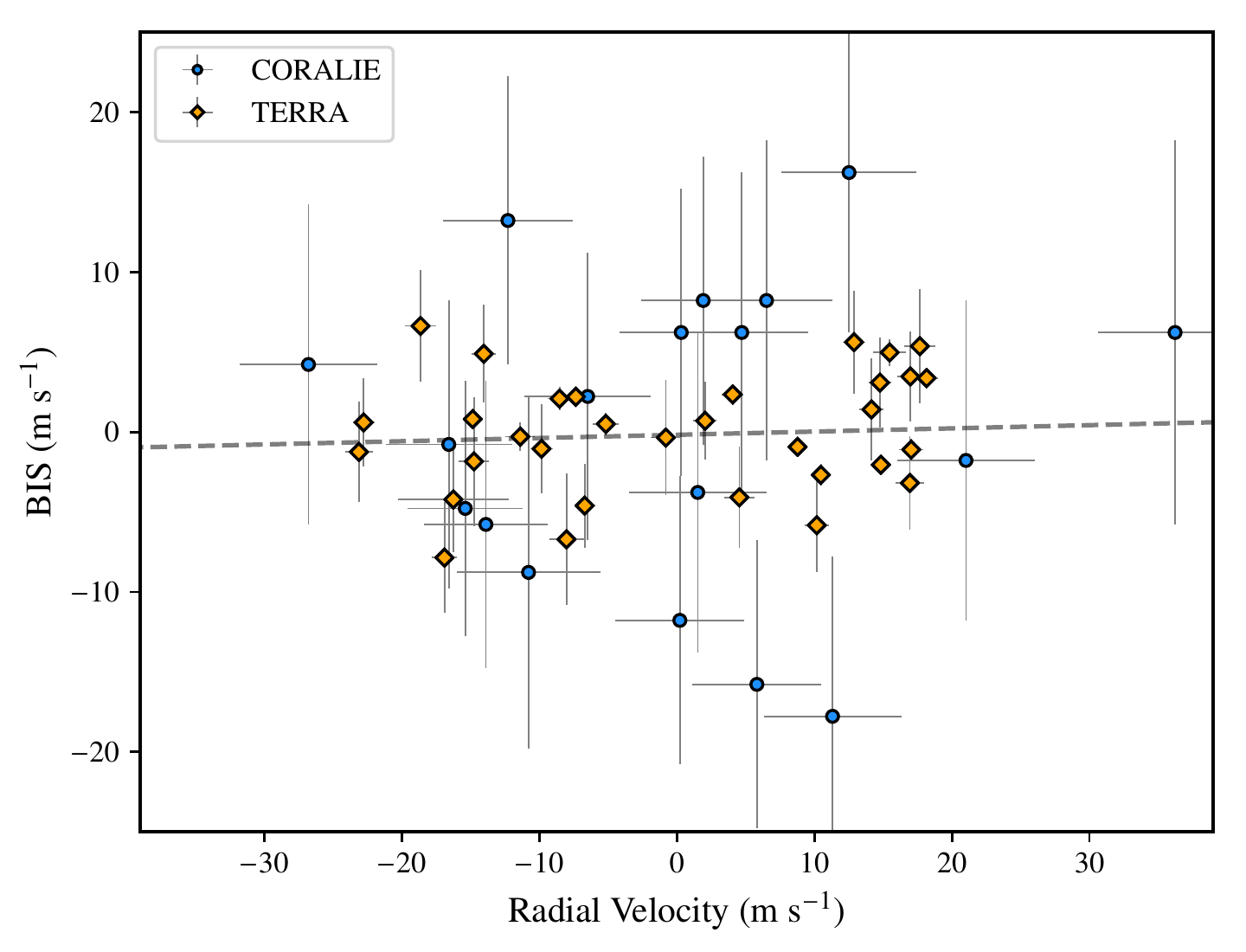}\label{fig:bis}
\caption{{\bf Spectral line bisector inverse slope measurements as a
  function of the radial-velocities.} The orange diamonds and blue
  circles relate to measurements made using HARPS and Coralie,
  respectively.  The best fit linear trend is shown by the dashed
  line, and a key in the upper left indicates the origin of the data
  points.
}
\end{figure}

\noindent{\bf Follow-up High-angular Resolution Imaging}

\underline{NIRC2 at Keck}

As part of our standard process for validating transiting exoplanets,
we observed \starname\, with infrared high-resolution adaptive optics
(AO) imaging at Keck Observatory \citep{ciardi15}.  The Keck Observatory observations were made with the NIRC2 instrument on Keck-II behind the natural guide star AO system.  The observations were made on UT 2018 Nov 22 following the standard 3-point dither pattern that is used with NIRC2 to avoid the left lower quadrant of the detector which is typically noisier than the other three quadrants. The dither pattern step size was $3''$ and was repeated twice, with each dither offset from the previous one by $0.5''$.  

The observations were made in the narrow-band $Br-\gamma$ filter
$(\lambda_o = 2.1686; \Delta\lambda = 0.0326\mu$m) with an integration
time of 2 seconds with one coadd per frame for a total of 18 seconds on target.  The camera was in the narrow-angle mode with a full field of view of $\sim10''$ and a pixel scale of approximately $0.0099442''$ per pixel. The Keck AO observations show no additional stellar companions were detected to within a resolution $\sim 0.056''$ FWHM (Figure~8 left).

The sensitivities of the final combined AO image were determined by
injecting simulated sources azimuthally around the primary target
every $45^\circ $ at separations of integer multiples of the central
source's FWHM \citep{furlan17}. The brightness of each injected source
was scaled until standard aperture photometry detected it with
$5\sigma$ significance. The resulting brightness of the injected
sources relative to the target set the contrast limits at that
injection location. The final $5\sigma$ limit at each separation was
determined from the average of all of the determined limits at that
separation and the uncertainty on the limit was set by the rms
dispersion of the azimuthal slices at a given radial distance. The
sensitivity curve is shown in the left panel of Figure~8, along with
an inset image zoomed to primary target showing no other companion
stars. \\

\begin{figure}
\includegraphics[width=\textwidth]{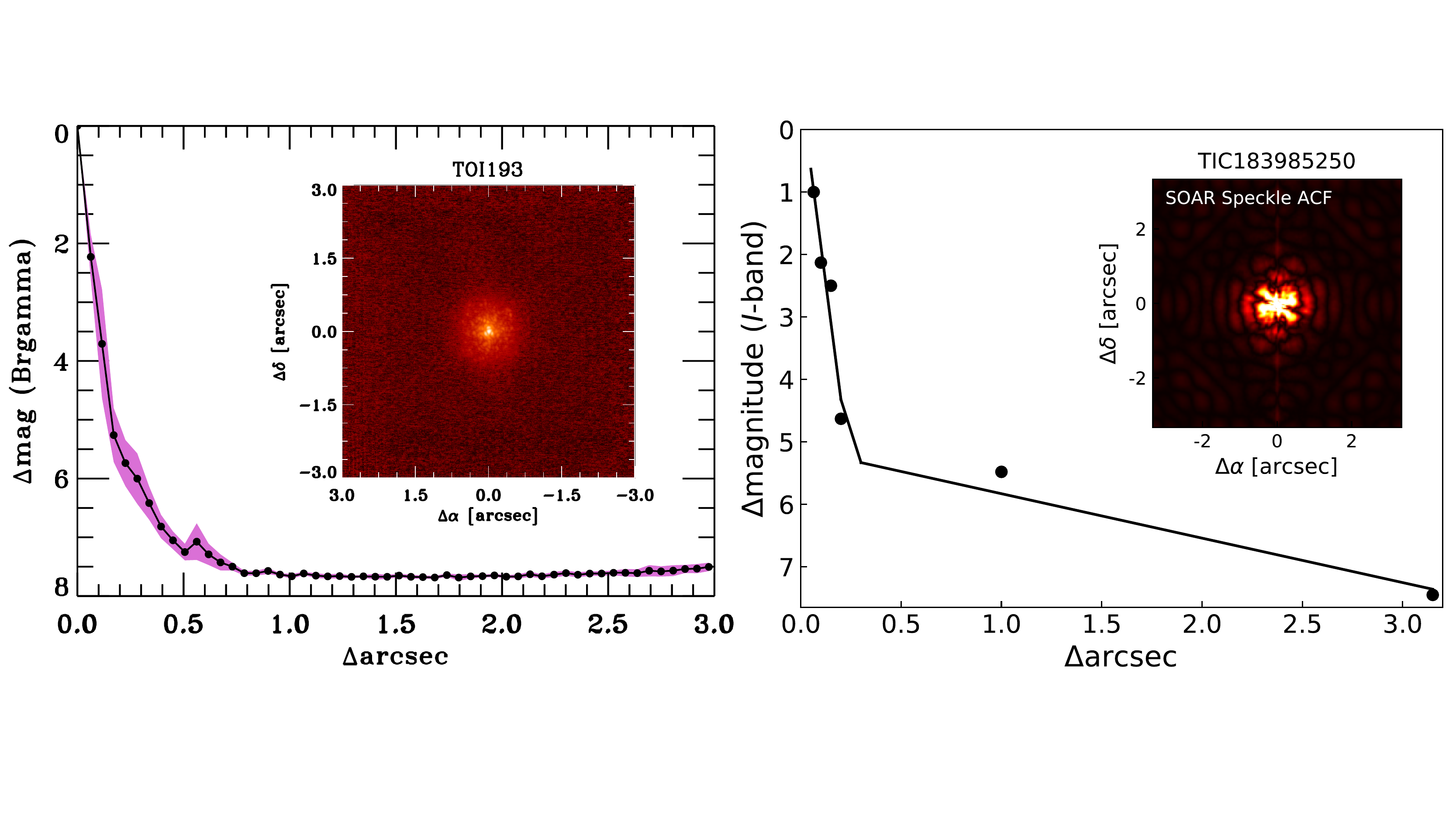}\label{fig:ao}
\caption{{\bf Companion sensitivity for the Keck NIRC2 adaptive optics imaging
  and the SOAR Adaptive Optics Module (SAM).}  
  For NIRC2 (left), the black points represent the 5$\sigma$ limits and are
  separated in steps of 1 FWHM ($\sim 0.05''$); the purple represents
  the azimuthal dispersion (1$\sigma$) of the contrast determinations
  (see text). The inset image is of the primary target showing no
  additional companions within 3$''$ of the target.  For SAM (right) the black
curve also represents the 5$\sigma$ limit, and the black data points
mark the sampling.  The inset also shows the speckle image of the
star, constructed from the Auto-Correlation Function.}
\end{figure}

\underline{HRCam at SOAR}

In addition to the Keck observations, we also searched for nearby
sources to \starname\, with SOuthern Astrophysical Research (SOAR) speckle imaging on 21 December 2018
UT, using the high resolution camera (HRCam) imager. Observations were performed in the $I$-band, which is a similar
visible bandpass to that of \tess.  Observations consisted of 400
frames, consisting of a 200$\times$200 binned pixels region of
interest, centered on the star.  Each individual frame is 6.3$''$ on a
side, with a pixel  scale  of  0.01575$''$  and  2$\times$2  binning,
with an observation time of $\sim$11 s, and using an Andor iXon-888
camera.  More details of the observations and processing are
available in \citet{ziegler19}. 

The 5$\sigma$ contrast curve and speckle auto-correlation function image are
shown in the right panel of Figure~8.  No nearby sources were
detected within 3\arcsec of \starname, down to a contrast limit of
6$\--$7 magnitudes in the $I$-band.  We can also rule out brighter
background blends very close to the star, down to around 0.1$''$
separation.  Combining the results from Keck and SOAR, we can be rule
out background blended eclipsing binaries contaminating the \tess\,
large aperture used to build the \starname\, light curve.\\

\noindent{\bf Stellar Parameters}\label{star}

To calculate the stellar parameters for \starname\, we used four
different methods, with three of them applied to the three different
sets of spectra we obtained from NRES, TRES, and HARPS, and a
photometric method that used our new tool \ariadne.  For the NRES
spectra, we used the combination of SpecMatch and Gaia DR2 to perform
the spectral classification, following the procedures explained in
\citet{fulton18}. TRES spectral observations used the Spectral
Parameter Classification (\spc; \citealp{buchhave12}) tool to
calculate the stellar parameters, whereas we used the Spectroscopic
Parameters and atmosphEric ChemIstriEs of Stars (\species;
\citealp{soto18}) and the Zonal Atmospheric Stellar Parameters
Estimator (\zaspe; \citealp{brahm17a}) algorithms to analyse the HARPS
spectra.  Details of these methods can be found in each of the listed
publications, yet in brief, \spc\, and \zaspe\, calculate the
parameters by comparing the spectra to Kurucz synthetic model grids \citep{kurucz92}, either by direct spectral fitting, or by cross correlation.  In this way, regions of the spectra that are sensitive to changes in stellar parameters can allow parameters to be estimated by searching for the best matching spectral model.  

On the other hand, \species\, uses an automatic approach to calculate
equivalent widths for large numbers of atomic spectral lines of
interest, Fe\sc~i\rm\, for instance.  The code then calculates the
radiative transfer equation using \moog\ \citep{sneden73}, applying
ATLAS9 model atmospheres \citep{castelli04}, and converges on the stellar parameters using an iterative line rejection procedure.  Convergence is reached once the constraints of having no statistical trend between abundances calculated from Fe\sc~i\rm\, and Fe\sc~ii\rm\, for example, reaches a pre-determined threshold value.  

Each of these three methods return consistent results for the majority
of the bulk parameters, in particular the stellar effective
temperature is in excellent agreement, with a mean value of
5480$\pm$42~K, along with the surface gravity (\logg) of the star,
which is found to be 4.47$\pm$0.11~dex.  For the metallicity of the
star, all three methods find the star to be metal-rich, with a mean
value of +0.27$\pm$0.04~dex.  For the main parameters of interest in
this work, the stellar mass and radius, we used two different methods,
with the mass value coming from the combination of the GAIA DR2
parallax for the star \citep{gaia16,gaia18}, along with either the
MESA Isochrones and Stellar Tracks (MIST; \citealp{dotter16}) models,
or the Yonsei-Yale (YY; \citealp{yi01}) isochrones, and we find a value of 1.02$^{+0.02}_{-0.03}$~\msun.  

For the radius, we used the \ariadne\ code \citep{vines2020}, which is
a new python tool designed to automatically fit stellar spectral
energy distributions in a Bayesian Model Averaging framework.  We
convolved Phoenix v2 \citep{husser13}, BT-Settl, BT-Cond
\citep{allard12}, BT-NextGen \citep{hauschildt99,allard12},
\citet{castelli04}, and \citet{kurucz93} model grids with commonly
available filter bandpasses: UBVRI; 2MASS JHK$_s$; SDSS ugriz;
ALL-WISE W1 and W2; Gaia G, RP, and BP; Pan-STARRS griwyz; Stromgren
uvby; GALEX NUV and FUV; Spitzer/IRAC 3.6$\mu$m and 4.5$\mu$m; TESS;
Kepler; and NGTS creating six different model grids, which we then
interpolated in Teff$\--$log~g$\--$[Fe/H] space. ARIADNE also fits for
the radius, distance, Av and excess noise terms for each photometry
point used, in order to account for possible underestimated
uncertainties. We used the \species\, results as priors for the \teff,
\logg, and [Fe/H], and the distance is constrained by the Gaia DR2
parallax, after correcting it by the offset found by
\citet{stassun18}.  The radius has a prior based on GAIA’s radius
estimate and the Av has a flat prior limited to 0.029, as per the
re-calibrated SFD galaxy dust map \citep{schlegel98,schlafly11}. We
performed the fit using dynesty’s nested sampler \citep{speagle20}, which returns the Bayesian evidence of each model, and then afterwards we averaged each model posterior samples weighted by their respective normalized evidence.  This returned a final stellar radius of 0.949$\pm$0.006~\rsun.

As \planet\, appears as an odd-ball when scrutinising its mass and radius, we want to be sure that the stellar radius is not biased in the sense that the star is really an evolved star, much larger than the stellar modelling predicts, and hence the planet is more likely a UHJ.  Although we have arrived at the same values from three different analyses and instrumental data sets, we can add more confidence to the results by studying the stellar density throughout the MCMC modelling process, when assuming the planet's orbit is circular.  In this case, we place a log-uniform prior on the stellar density, constrained to be within 100$\--$10'000~kgm$^{-3}$, and then study how it changes as a function of $\rpl/\rstar$.  

We find that the distribution is bimodal (Figure~9), with the most likely stellar density region given by the lower, more densely constrained part of the parameter space in the figure.  The upper mode in the figure, pushing towards higher stellar densities and lower values of $\rpl/\rstar$, is arguing towards the star being an M dwarf, which is ruled out by the high resolution spectroscopic data, and is inconsistent with our global-modelling effort (less probable part of the posterior space).  This mode is also only consistent with a very narrow set of limb-darkening coefficients, all of which are inconsistent at several sigma with theoretical models, whereas the lower, more probable mode, has a wide range of possible limb-darkening coefficients, which are all in agreement with theoretical models.  Therefore, this test rules out a more evolved state for the star in either case, with the higher probability mode being in excellent agreement with the results from the stellar modelling.

\begin{figure}
\includegraphics[width=0.8\textwidth]{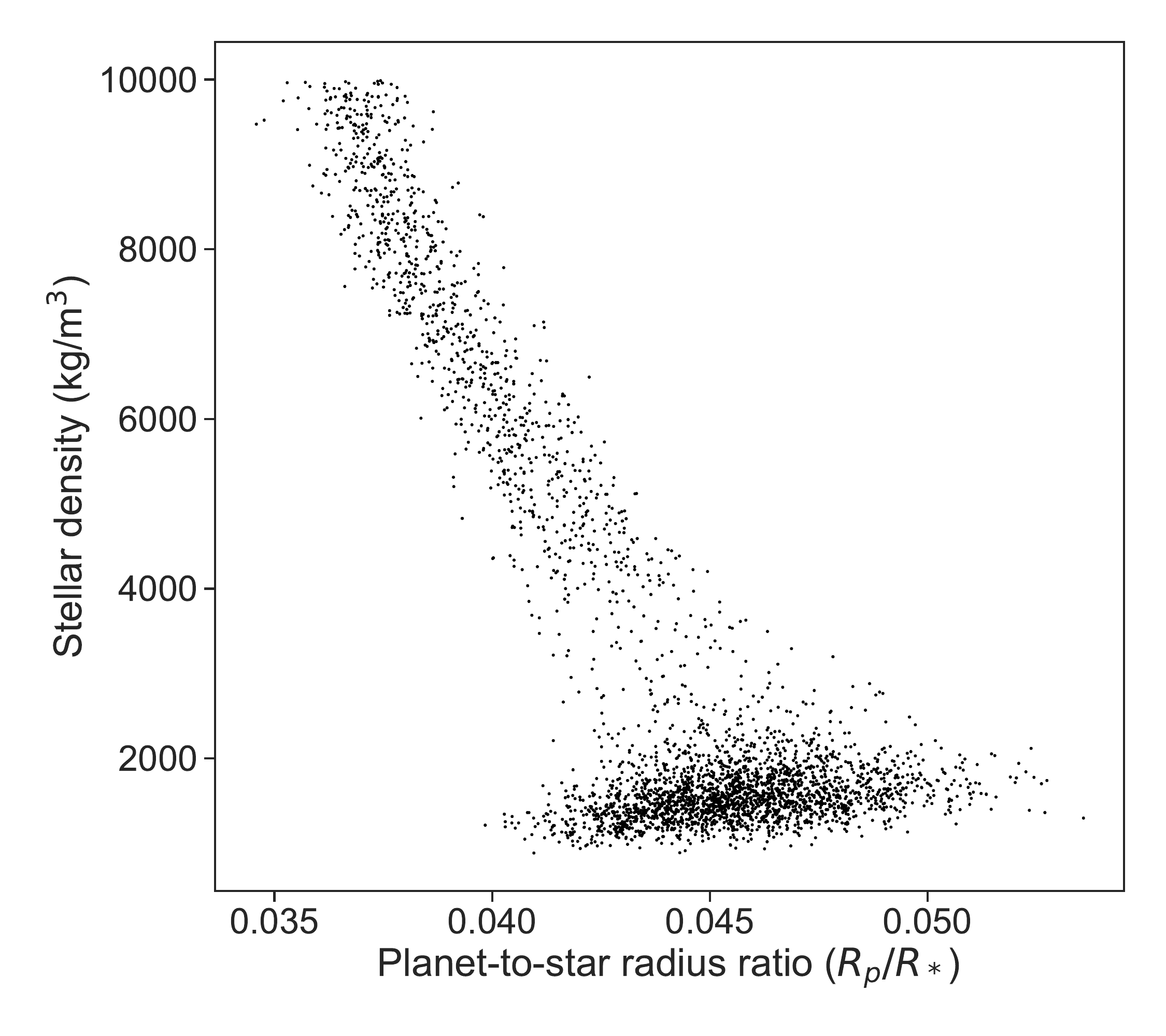}\label{fig:rprs}
\caption{{\bf Stellar density as a function of $\rpl/\rstar$} when modelling the \tess\, light curve with a log-uniform prior on the stellar density and the planetary eccentricity constrained to be zero.}
\end{figure}

Finally, for the confirmation of the transit and radial-velocity
parameters it is prudent to analyse the activity of the star, in order
to assess the impact that any activity could have on the measurements.
From the above analyses we find the star to be a very slow rotator,
with a HARPS \vsini\, limit of 1.06$\pm$0.37~\kms, lower than the
projected solar \vsini\, value (1.6$\pm$0.3~\kms) determined from
HARPS spectral analysis \citep{pavlenko12}, indicating a slowly
rotating, and therefore inactive star.  Given the calculated radius of
the star, such a slow rotation gives rise to an upper limit of the
rotation period to be 45~d.  If the planetary orbit is aligned with
the stellar plane of rotation, such that we can assume the inclination
angle is the same, then this value is the absolute rotation period.
Kepler Space Telescope data analysis of old field stars of this
spectral type, have rotation periods ranging from a few days for the
youngest stars, with a peak around 20~d, and a sharp fall after this
with a tail reaching up to almost 100~d \citep{mcquillan14}.  A
rotation period of 45~d would place \starname\, in the upper tail of
the Kepler distribution, indicating the star is old, and agreeing with
the combined age estimate of 2.0$^{+1.3}_{-0.9}$~Gyrs.  This result
would also suggest that the activity of the star should be weak.  We
calculate the activity using the Ca\sc~ii\rm\, HK lines, following the
analysis procedures and methods presented in
\citet{JenkinsEtal2006mnrasActivityCatalogue,JenkinsEtal2008aaMetallicitiesActivities,jenkins11,jenkins17}.
We find the star to be inactive, with a HARPS $S$-index of
0.148$\pm$0.008, which relates to a mean \logrhk\, of
-5.10$\pm$0.04~dex.  Gyrochronology relations \citep{mamajek08} would therefore suggest an age closer to $\sim$5~Gyrs or so, again confirming that the star should not be young.  Taken all together, \starname\, can be classed as an inactive and metal-rich solar analogue star, and all key properties can be found in Table~\ref{tab:star}.
\\

\noindent{\bf Global Modelling}\label{globmod}

As stated in the main text, the global modeling of the data was 
performed using \exonailer\ \citep{juliet}. This code uses
\texttt{batman} \citep{batman} to model the transit lightcurves and
\texttt{radvel} \citep{radvel} to model the 
radial-velocities. We performed the posterior sampling using 
MultiNest \citep{MultiNest} via the PyMultiNest wrapper \citep{PyMultiNest}.

The fit was parameterized by the parameters $r_1$ and $r_2$, both 
having uniform distributions between 0 and 1, which 
are transformations of the planet-to-star radius ratio $p$ and 
impact parameter $b$ that allow an efficient exploration of the 
parameter space \citep{espinozaEBP}. In addition, we fitted for the 
stellar density by assuming a prior given by the value obtained with 
our analysis of the stellar properties, assuming a normal prior for 
this parameter with a mean of 1810 kg/m$^3$ and standard deviation of 
130 kg/m$^3$. We parameterized the limb-darkening effect using a quadratic 
law defined by parameters $u_1$ and $u_2$; however, we use an 
uninformative parameterization scheme \citep{kipping:ld} in which we fit 
for $q_1 = (u_1 + u_2)^2$ and $q_2 = u_1/(2u_1+2u_2)$ with 
$q_1$ and $q_2$ having uniform priors between 0 and 1. For the 
radial-velocity parameters, we used wide priors for both the systemic 
radial-velocity of each instrument and the possible jitter terms, 
added in quadrature to the data. 

For the photometry, we considered unitary 
dilution factors for the \tess\, NGTS and LCOGT photometry after leaving 
them as free parameters and observing that it was not needed based 
on the posterior evidence of the fits. This is consistent with 
the a-priori knowledge that the only source detected by Gaia DR2 
within the \tess\, aperture is a couple of faint sources to the 
south-east of the target, the brighter of which has $\Delta G = 5.4$ 
with the target. If we assume the Gaia passband to be similar to the 
\tess\, passband, this would imply a dilution factor $D > 0.99$, which is negligible for our purposes. For the \tess\ photometry, no extra noise model nor jitter term was needed to be added according to the bayesian evidence of fits incorporating those extra terms. For the NGTS observations, we considered the data of the target from the nine different telescopes as independent photometric datasets (i.e., having 
independent out-of-transit baseline fluxes in the joint fit), that 
share the same limb-darkening coefficients. We initially added 
photometric jitter terms to all the NGTS observations, but found that fits without them for all 
instruments were preferred by looking at the bayesian evidences 
of both fits. For the LCOGT data, we used gaussian process in time to
detrend a smooth trend observed in the data. A kernel which was a
product of an exponential and a matern 3/2 was used, and a jitter term
was also fitted and added in quadrature to the reported uncertainties
in the data --- this was the model that showed the largest bayesian
evidence. We note that fitting the lightcurves independently
  provides statistically similar transit depths to the joint model,
  showing that all are in statistical agreement. Finally, an eccentric orbit is ruled out by our 
data with an odds ratio of 49:1 in favor of a circular orbit; 
the eccentric fit, performed by parameterizing the eccentricity 
and argument of periastron via $\mathcal{S}_1 = \sqrt{e}\cos \omega$ and 
$\mathcal{S}_2 = \sqrt{e}\sin \omega$, gives an eccentricity given 
our data of $e<0.058$ with a 95\% credibility.

With all the photometry in hand, we could also compare individually each light curve transit model to test if they are in statistical agreement, or any biases exist, such that the radius measurement is biased.  We proceeded to again fit each light curve independently with \exonailer, recording the transit model depths to test for statistical differences.  As expected, we found the \tess\, photometry produced the most precise value ($T_{d,\tess}$ = 2299$^{+320}_{-240}$~ppm), with the LCO and NGTS fits arriving at values of $T_{d,LCO}$ = 1925$^{+620}_{-400}$~ppm and 1594$^{+980}_{-715}$~ppm, respectively.  All three are in statistical agreement.  We also jointly modeled the LCO and NGTS lightcurves to provide a more constrained comparison with the \tess\, photometry, and found a value of $T_{d,LCO+NGTS}$ = 1678$^{+540}_{-290}$~ppm, again in statistical agreement with the \tess\, value.  Therefore, we can be confident that all three instruments provide a similar description for the planet's physical size.
\\

\noindent{\bf Transit Timing Variations}\label{ttvs}

The Transit Timing Variations (TTVs) of \planet\, was measured using
the \exofast\ \citep{eastman2013,eastman2017} code. \exofast\, uses the Differential Evolution Markov chain Monte Carlo method (DE-MC) to derive the values and their uncertainties of the stellar, orbital and physical parameters of the system. For the TTV analysis of \planet\, we fixed the stellar and orbital parameters to the values obtained from the global fit performed by \species\, and \exonailer\,, except for the transit time of each light curve and their baseline flux.  

In a Keplerian orbit, the transit time of an exoplanet follows a linear function of the transit epoch number (E):
\begin{equation}
   T_{c}(E) = T_{c}(0) + PE 
\end{equation}

Where P is the orbital period of the exoplanet and $T_{c}(0)$ is the optimal transit time in a arbitrary zero epoch and corresponds to the time that is least covariant with the period and has the smallest uncertainty. Our best-fitted value from \exofast\, is: $T_{c}(0)\,=\,2458354.2145\pm0.0012$\,BJD.

All the transit times were allowed to move from the linear ephemeris and each one was considered as one independent TTV parameter in the \exofast\,'s fitting, resulting in 33 parameters to fit. The best fit results are shown in Figure~10, where the grey area corresponds to the 1$\sigma$ of the linear ephemeris shown in Equation (1). 

We found no evidence of a clear periodic variation in the transit time. The RMS variation from the linear ephemeris is $\sigma = 181.8\ $sec. There are only two values above the 2$\sigma$ limit, if we remove them the RMS deviation is reduced to 155.9 sec. On the other hand, the reduced chi-squared is $\chi^2_{red}=1.23$, which is an indicator that the transit times fit accordingly with the proposed linear ephemeris.

In conclusion, the existence of transit timing variations in \planet\, is not evident for the time-span of our transit data. In addition, with the apparent lack of another short period signal in the RV data, this suggest that there is no other inner companion in the planetary system.  Any other tertiary companion must be far from \planet, such that the gravitation or tidal interactions are small, and the linear trend in the RVs might be pointing in that direction.\\

\begin{figure}
\includegraphics[width=0.8\textwidth]{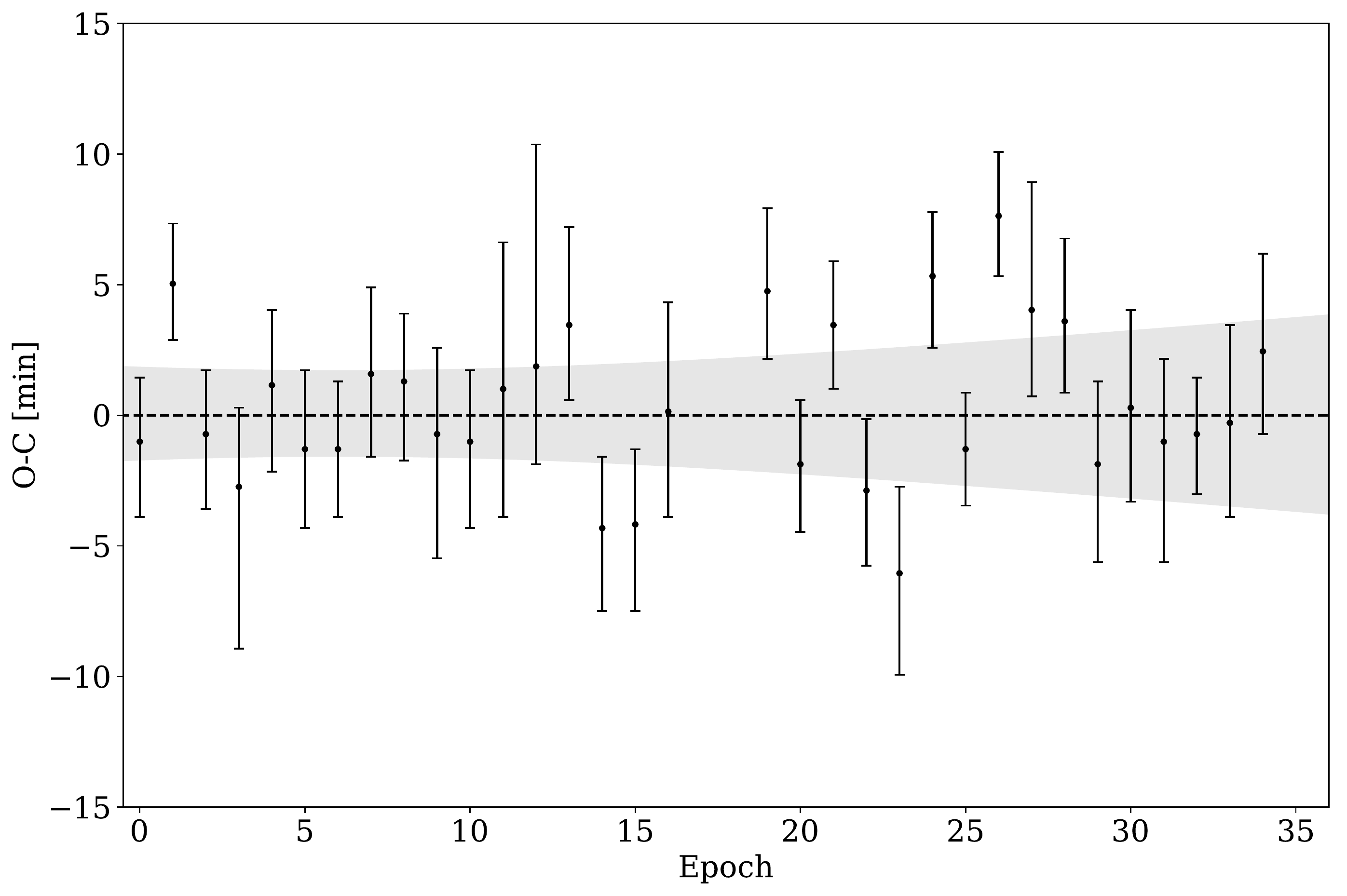}\label{fig:ttv}
\caption{{\bf Observed minus computed mid-transit times of \planet.} The residuals (TTV) of the transit times are shown considering the proposed linear ephemeris. The dashed line corresponds to zero variation and the grey area is the propagation of 1$\sigma$ uncertainties, considering the optimal transit time from \exofast\, and the period from \exonailer\,. The epoch 0 is the first light curve obtained by \tess\, and is also the corresponding epoch of the optimal transit time. The TTV values shown in this plot fit accordingly with the proposed linear ephemeris ($\chi^2_{red}=1.23$).}

\end{figure}

\noindent{\bf Metallicity Analysis}\label{met_analysis}

The correlation between the presence of giant planets and host star
metallicity has been well established
\citep{gonzalez97,fischer05,jenkins17,maldonado18}, along with the
apparent lack of any correlation for smaller planets
\citep{jenkins13,buchhave12,buchhave14}.  We studied the small sample
of known USP planets and Ultra Hot Jupiters (UHJs, the gas giant
planets with orbit periods of less than 1~day), using values taken
from the TEPCat database \citep{southworth11}, whilst recalculating
metallicities for those where we could find their spectra, ($\sim$half
the sample), using \species\ (\citeauthor{soto18}).  We found a similar general trend, whereby the USP planets tend to orbit more metal-poor stars when compared with the UHJs, however the sample is small enough that single outliers bias the statistics, therefore we extended slightly the orbital period selection out to 1.3 days, increasing the sample by over 55\%.  With this updated sample, we find a Kolmogorov-Smirnov (KS) test probability of only 1\% that the USP planets and UHJs are drawn from the same parent population.  

A couple of notable exceptions to the trend here are the planets 55~Cancri~e and WASP-47~e, both small USP planets that orbit very metal-rich stars.  However, there exists additional gas giant planets in these systems, meaning they still follow the overall picture.  If we exclude these two, the KS probability drops to 0.1\% that the populations are statistically similar.  The diversity of USP planets is high, therefore many more detections are needed to statistically constrain the populations in this respect.  We also require more UHJs to build up a statistical sample, since the subsolar metallicity of WASP-43 can also bias the tests.  If we look at the density-metallicity parameter space (Figure~4), there are indications of a general trend whereby the low-density planets are mostly UHJs orbiting metal-rich stars, and the higher density USP planets orbit more metal-poor stars.

\subsection{Data availability}

Photometric data that support the findings of this study are
publically available from the Mikulski Archive for Space Telescopes
(MAST; http://archive.stsci.edu/) under the TESS Mission link.
All radial-velocity data re available from the corresponding author 
upon reasonable request. Raw and processed spectra can be obtained 
from the European Southern Observatory’s data archive at 
http://archive.eso.org.

\subsection{Code Availability}

All codes necessary for the reproduction of this work are publically
available through the GitHub repository, as follows: \\
{\bf EMPEROR:} https://github.com/ReddTea/astroEMPEROR \\
{\bf Juliet:} https://github.com/nespinoza/juliet \\
{\bf SPECIES:} https://github.com/msotov/SPECIES \\
{\bf ARIADNE:} https://www.github.com/jvines/astroARIADNE \\
{\bf CERES:} https://github.com/rabrahm/ceres \\
{\bf ZASPE} https://github.com/rabrahm/zaspe \\

\bibliography{refs}
\bibliographystyle{mn2e_v1}

\end{document}